\documentclass[]{interact}
\usepackage{epstopdf}

\usepackage[numbers,sort]{natbib} 

\renewcommand\bibfont{\fontsize{10}{12}\selectfont}

\usepackage[utf8]{inputenc}
\usepackage{textcomp}
\usepackage{graphicx}
\usepackage{amsmath}
\usepackage[version=4]{mhchem}
\usepackage{siunitx}
\usepackage{longtable,tabularx}
\title{Improved pressure-gradient sensor for the prediction of separation onset in RANS models}

\author{
\name{Kevin Patrick Griffin,\thanks{CONTACT K.~P. Griffin. Email: kgriffin@nrel.gov}  Ganesh Vijayakumar, Ashesh Sharma, and Michael A. Sprague}
\affil{National Renewable Energy Laboratory, Golden, CO, 80401.}
}

\usepackage{subcaption}
\usepackage{floatrow}
\usepackage{todonotes}
\newcommand{\pder}[2]{\frac{\partial#1}{\partial#2}} 
\newcommand{\pderi}[2]{\partial#1 / \partial#2} 

\usepackage{glossaries}
\newacronym{dns}{DNS}{direct numerical simulation}
\newacronym{rans}{RANS}{Reynolds-averaged Navier-Stokes}
\newacronym{les}{LES}{large-eddy simulation}
\newacronym{apg}{APG}{adverse pressure gradient}
\newacronym{fpg}{FPG}{favorable pressure gradient}
\newacronym{zpg}{ZPG}{zero pressure gradient}
\newacronym{bl}{BL}{boundary layer}
\newacronym{sst}{SST}{shear stress transport}
\newacronym{nrel}{NREL}{National Renewable Energy Laboratory}

\makeglossaries
\glsdisablehyper
\glsenableentrycount

\begin{document}

\maketitle

\begin{abstract}
We improve upon two key aspects of the Menter shear stress transport (SST) turbulence model: (1) We propose a more robust adverse pressure gradient sensor based on the strength of the pressure gradient in the direction of the local mean flow; (2) We propose two alternative eddy viscosity models to be used in the adverse pressure gradient regions identified by our sensor. Direct numerical simulations of the Boeing Gaussian bump are used to identify the terms in the baseline SST model that need correction, and \textit{a posteriori} Reynolds-averaged Navier-Stokes calculations are used to calibrate coefficient values, leading to a model that is both physics driven and data informed. The two sensor-equipped models are applied to two thick airfoils representative of modern wind turbine applications, the FFA-W3-301 and the DU00-W-212. The proposed models improve the prediction of stall (onset of separation) with respect to the prediction of the baseline SST model. 
\end{abstract}



\textbf{Keywords-}
k-omega SST

\section{Introduction} \label{sec:intro}
The prediction of separation in turbulent boundary layers is of paramount importance for aerodynamic design but remains a challenge, even with state-of-the-art \gls{rans} modeling. In this work, we focus on the 2003 Menter \gls{sst} model \citep{Menter2003}  (hereafter refered to as the baseline SST model). Despite being developed two decades ago, it remains a state-of-the-art open-source model for many research and industrial applications \citep{Menter2009,Menter2021}, including the prediction of stall in aircraft and wind turbines. Nonetheless, it cannot reliably predict separation in flows over smooth bodies, which motivates the present investigation of ways to improve the model. The remainder of this section details the equations of the baseline SST model, discusses existing approaches to modify these equations to improve the prediction of separation, the impact of these modifications on \gls{rans} predictions, and finally outlines criteria for model evaluation to guide model development in subsequent sections.

\subsection{The baseline SST model}
In this section, for completeness, we state the details of the 2003 version of the SST model as given in \cite{Menter2003}. The model consists of a transport equation for $k$, the turbulent kinetic energy,
\begin{equation} \label{eq:k}
    \pder{\rho k}{t} + \pder{\rho u_j k}{x_j} = P - \beta^* \rho \omega k + \pder{}{x_j}[(\mu + \sigma_k \mu_t) \pder{k}{x_j}],
\end{equation}
and a transport equation for $\omega$, the specific dissipation rate of kinetic energy,
\begin{equation} \label{eq:omega}
    \pder{\rho \omega}{t} + \pder{\rho u_j \omega}{x_j} = \frac{\gamma \rho }{\mu_t}P - \beta \rho \omega^2 + \pder{}{x_j}[(\mu + \sigma_\omega \mu_t) \pder{\omega}{x_j}] + 2(1-F_1) \frac{\rho \sigma_{\omega 2}}{\omega} \pder{k}{x_j} \pder{\omega}{x_j},
\end{equation}
where $\rho$ is the density, $u_j$ are the components of the (mean) velocity, $\mu$ is the viscosity, $\mu_t$ is the eddy (turbulent) viscosity, and $\beta^*=0.09$. The production of $k$ is defined as $P = \min(\tau_{ij} \pderi{u_i}{x_j},10\beta^*\rho\omega k)$, where the Reynolds shear stress tensor is modeled using the eddy viscosity assumption $\tau_{ij}=2\mu_t[S_{ij}-(\pderi{u_k}{x_k})\delta_{ij}/3]-2\rho k \delta_{ij}/3$, and the rate of strain tensor is defined as $S_{ij}=[\pderi{u_i}{x_j}+\pderi{u_j}{x_i}]/2$. 
The remaining model coefficients are determined from a mixing rule based on the $F_1$ boundary layer sensor as
\begin{equation} \label{eq:phi}
  \phi = F_1 \phi_1 + (1-F_1) \phi_2,
\end{equation}
where $\phi$ represents $\beta$, $\sigma_k$, $\sigma_\omega$, and $\gamma$, and the constants $\phi_1$ and $\phi_2$ are given as $\beta_1=0.075$, $\beta_2=0.0828$, $\gamma_1=5/9$, $\gamma_2=0.44$, $\sigma_{k1}=0.85$, $\sigma_{k2}=1.0$, $\sigma_{\omega 1}=0.5$, and $\sigma_{\omega 2}=0.856$.
$F_1$ is was designed to tend to unity inside boundary layers and to zero outside and is defined as
\begin{equation} \label{eq:F1}
  F_1 = \tanh(arg_1^4),
\end{equation} 
\begin{equation} \label{eq:arg1}
  arg_1 = \min \left[ \max \left( \frac{\sqrt{k}}{\beta^* \omega d}, \frac{500 \nu}{d^2 \omega} \right), \frac{4\rho \sigma_{\omega 2} k}{CD_{k \omega} d^2} \right],
\end{equation}
\begin{equation} \label{eq:CD}
  CD_{k \omega} = \max \left( 2 \rho \sigma_{\omega 2} \frac{1}{\omega} \pder{k}{x_j} \pder{\omega}{x_j}, 10^{-10} \right),
\end{equation}
where $d$ is the minimum wall distance.
The eddy viscosity is defined as
\begin{equation} \label{eq:mut}
  \mu_t = \frac{\rho a_1 k}{\max(a_1 \omega, S F_2)},
\end{equation}
where $a_1=0.31$, the strain rate magnitude $S = \sqrt{2 S_{ij} S_{ij}}$, and $F_2$ is a second boundary layer sensor designed to tend to unity inside boundary layers and zero outside. 
\begin{equation} \label{eq:F2}
  F_2 = \tanh(arg_2^2),
\end{equation}
\begin{equation} \label{eq:arg2}
  arg_2 = \max \left( 2\frac{\sqrt{k}}{\beta^* \omega d}, \frac{500 \nu}{d^2 \omega} \right).
\end{equation}
The model for $\mu_t$ was designed to recover $\mu_t = \rho k/\omega$ in most regions of the flow and switch to $\mu_t \sim \rho k/S$ in regions of strong \gls{apg}. This assumes that $a_1 \omega < SF_2$ is a good sensor for strong APG regions, which will be assessed in Section \ref{sec:model}.

\subsection{Performance of the SST and existing modifications}
\begin{figure}
\centering
\floatsetup[figure]{style=plaintop} 
\captionsetup[subfigure]{position=top, justification=raggedright, singlelinecheck=false} 
\subcaptionbox{}{\includegraphics[width=0.49\textwidth]{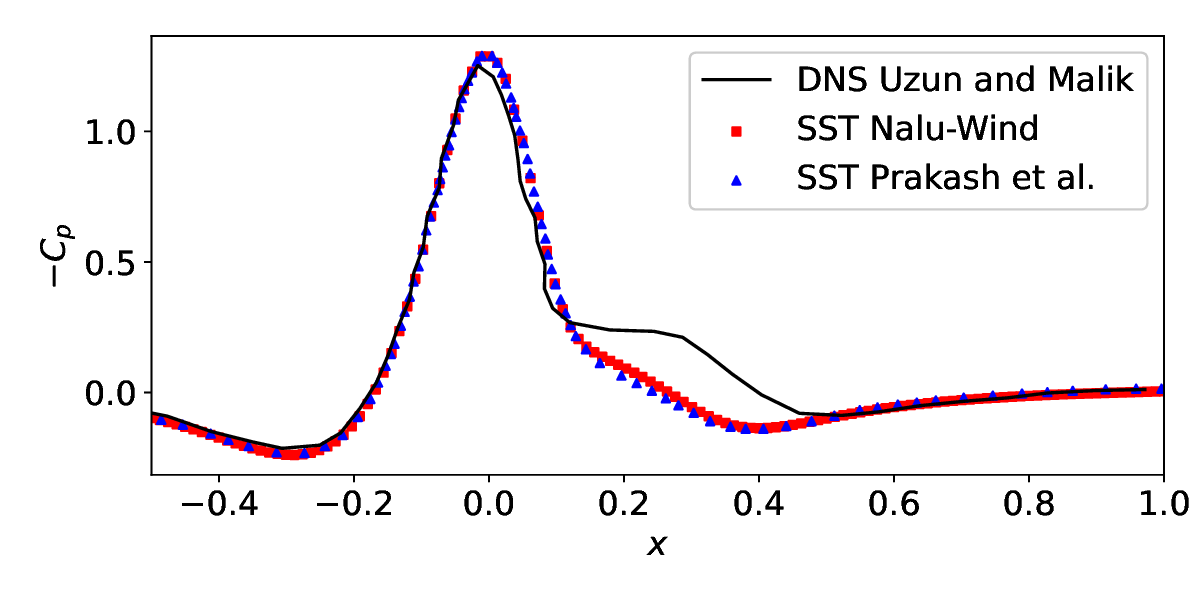}}
\subcaptionbox{}{\includegraphics[width=0.49\textwidth]{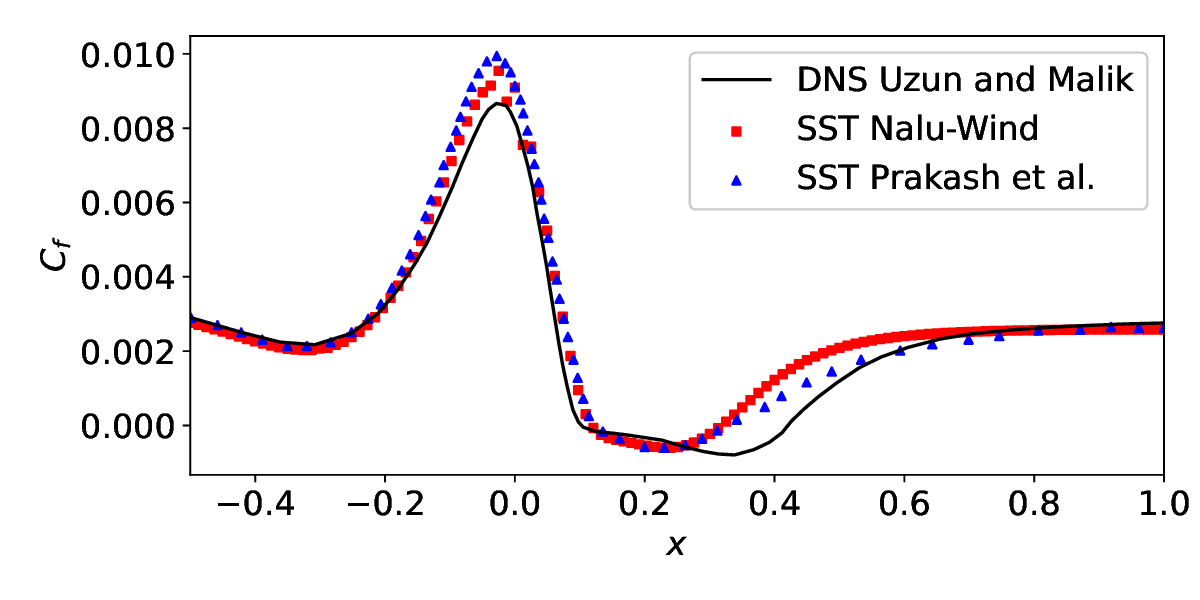}}
\caption{Coefficients of pressure (a) and friction (b) versus the streamwise coordinate for the flow over the Boeing Gaussian bump. The prediction by Nalu-Wind using the baseline SST RANS model is compared with the same model from \cite{Prakash2024} and \gls{dns} reference data from \cite{Uzun2022}.}
\label{fig:bump_baseline}
\end{figure}

\begin{figure} 
\centering 
\floatsetup[figure]{style=plaintop} 
\captionsetup[subfigure]{position=top, justification=raggedright, singlelinecheck=false} 
\subcaptionbox{}{\includegraphics[width=0.4\textwidth]{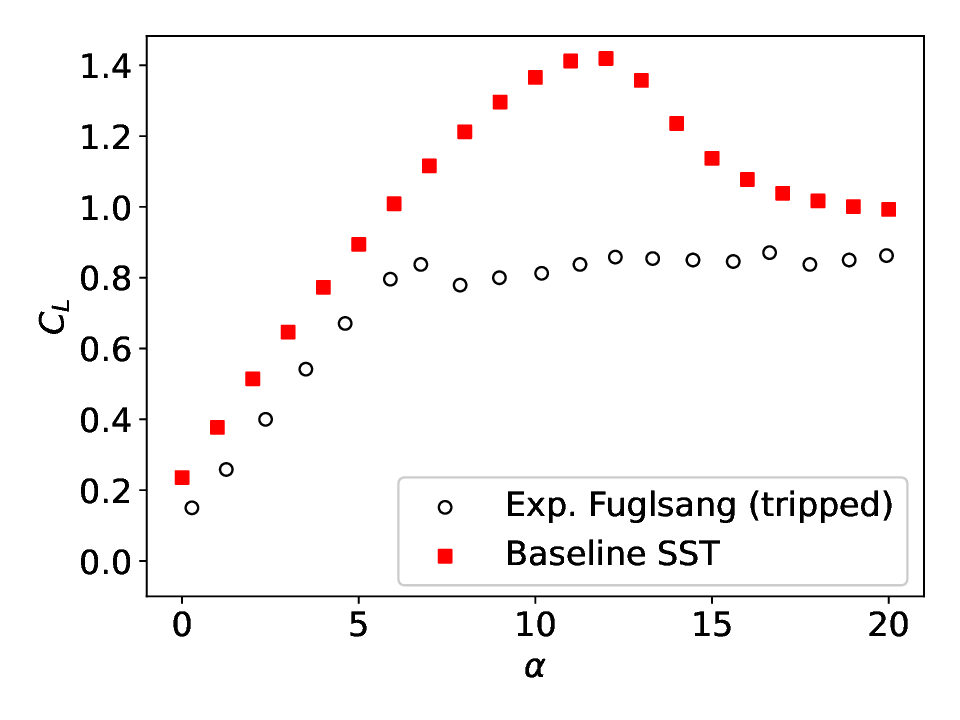}}
\subcaptionbox{}{\includegraphics[width=0.4\textwidth]{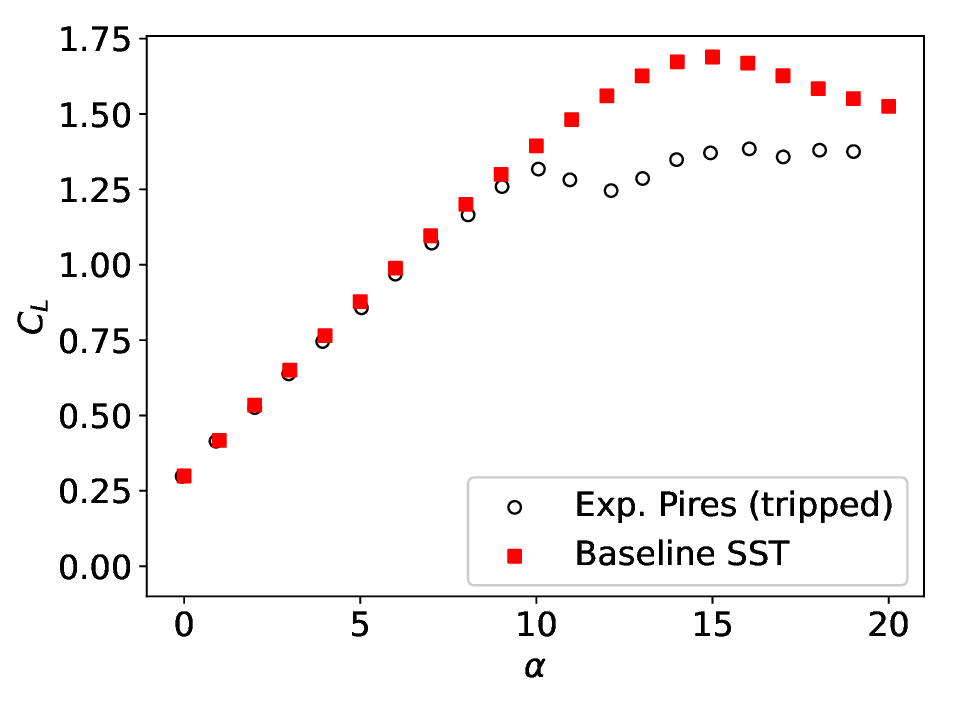}}
\caption{Lift coefficient predicted by the \gls{sst} model in Nalu-Wind (red) and the reference experiments \citep{Fuglsang1998,Pires2016} (black) for the FFA-W3-301 airfoil at $Re_c=1.6\times10^6$ (a) and DU00-W-212 airfoil at $Re_c=6\times10^6$ (b).}
\label{fig:airfoil_baseline}
\end{figure}

Although the SST model improves the prediction of separation compared to its constituent models (i.e., the model was constructed to recover the $k$-$\epsilon$ model in the far field and the $k$-$\omega$ model near walls) \citep{Menter1994}, deficiencies in predicting separation are widely reported in the literature\citep{Menter2003,Menter2021}. Here, we report two such examples using calculations with the Nalu-Wind CFD code; the code and simulation approaches are described in detail below. Figure \ref{fig:bump_baseline}a presents the coefficient of pressure $C_p=(p-p_\infty)/(\rho_\infty U_\infty^2/2)$ for the flow over the Boeing Gaussian bump, where $p$ is the pressure and the $\infty$ subscript indicates a quantity evaluated in the freestream. The \gls{dns} reference data \citep{Uzun2022} exhibit a flattening of the $C_p$ curve in the vicinity of $x/L=0.2$, which indicates the location of a separation bubble. Meanwhile, the \gls{sst} \gls{rans} data do not exhibit a flat region, indicating that the separation bubble is underpredicted. This also leads to the overprediction of the suction peak ($-C_p$ maximum). 
Likewise, the \gls{sst} model struggles to predict separation on a 2D FFA-W3-301 airfoil at a chord-based Reynolds number of $Re_c=1.6\times10^6$ and a DU00-W-212 airfoil at $Re_c=6\times10^6$. Figure \ref{fig:airfoil_baseline} displays the lift coefficient $C_L$ for the \gls{sst} and reference experimental data versus the angle of attack $\alpha$ reported in degrees. For the FFA-W3-301 airfoil, the stall (drop in $C_L$) is predicted by the \gls{sst} around $\alpha=12^\circ$, about $5^\circ$ later than in the experimental data of \cite{Fuglsang1998}. This behavior is similar to the under-separated prediction for the Boeing Gaussian bump and also helps motivate the study of the bump, a more canonical flow, for which detailed \gls{dns} statistics are available. It is also interesting to note that the \gls{sst} model slightly overpredicts the lift in the linear regime of $0<\alpha<7$ for the airfoil, which will be discussed in Section \ref{sec:results}. Similar overpredictions of lift for this flow have been observed in numerical calculations using XFOIL and EllipSys2D \citep{Fuglsang1998}.

\begin{figure}
\centering
\floatsetup[figure]{style=plaintop} 
\captionsetup[subfigure]{position=top, justification=raggedright, singlelinecheck=false} 
\subcaptionbox{}{\includegraphics[width=0.49\textwidth]{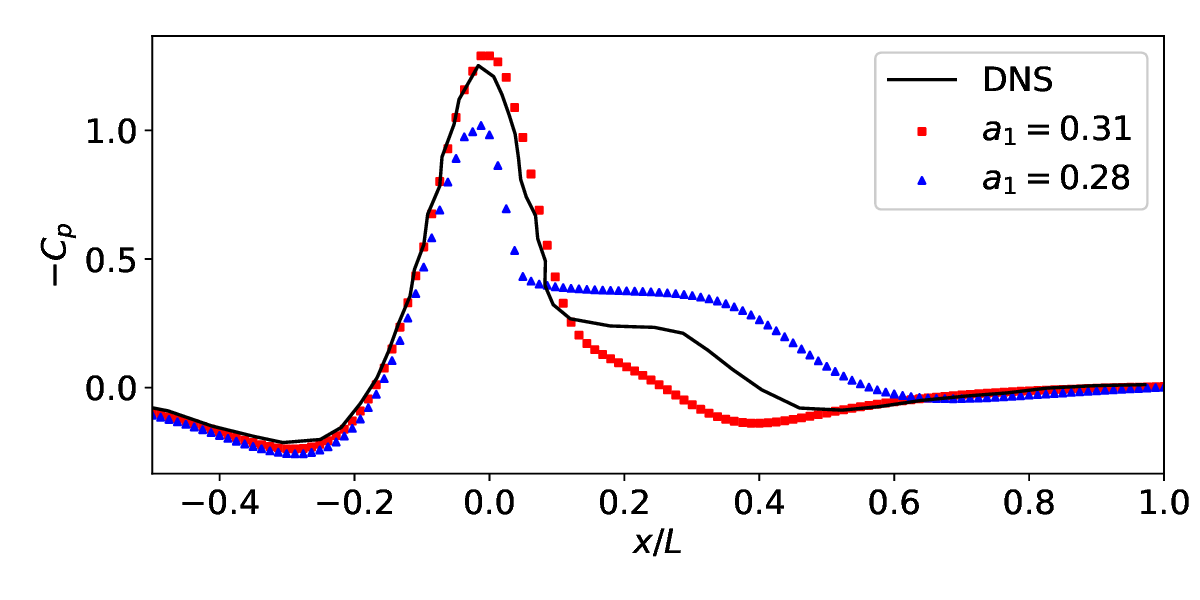}}
\subcaptionbox{}{\includegraphics[width=0.49\textwidth]{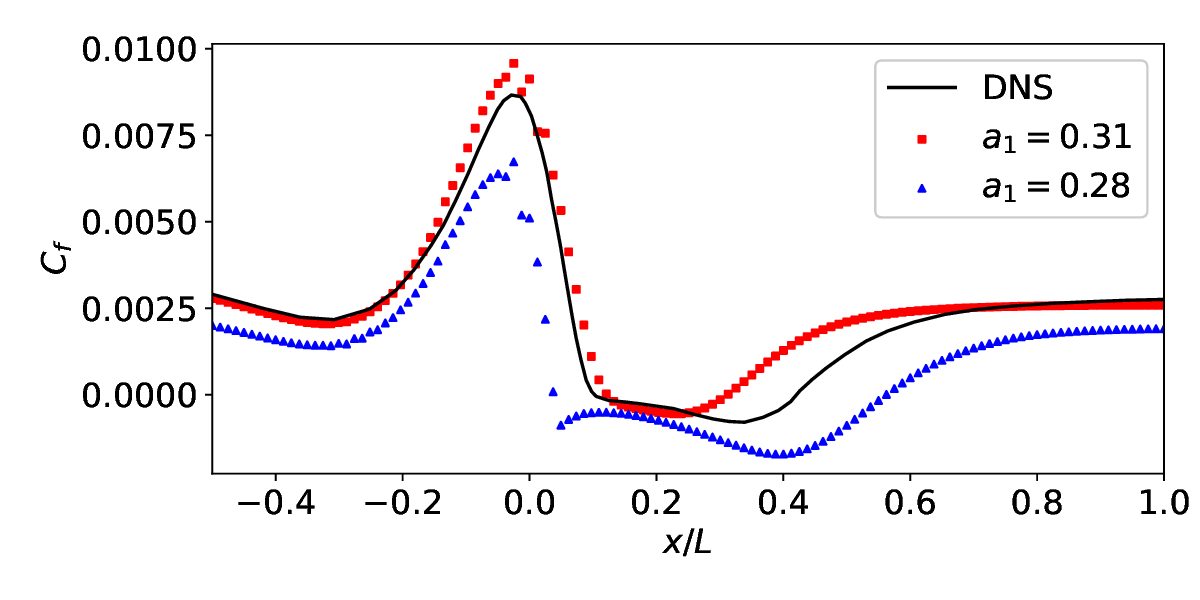}}
\caption{Effect of ad hoc tuning of $a_1$ in the \gls{sst} model in Nalu-Wind for the Boeing Gaussian bump is shown by results for the coefficients of pressure (a) and friction (b) with the baseline value $a_1=0.31$ in red and with the decreased value suggested by \cite{Matyushenko2016}  $a_1=0.28$ shown in blue. Reference \gls{dns} data \citep{Uzun2022} are shown in black.}
\label{fig:adhoc_a1}
\end{figure}

The challenge of predicting separation has motivated many investigations of data-informed approaches to turbulence modeling (see the review of \cite{Duraisamy2021}). Some efforts have focused on optimizing global coefficients in existing turbulence models (c.f., \citep{Barkalov2022}), while others have proposed methods of solving for spatially varying coefficients (c.f., \cite{Parish2016,Singh2017}). We will focus our discussion on three variations of the former that are specifically focused on the \gls{sst} model and that most directly shape our subsequent developments.
One such variant uses a modified value of $a_1$, the eddy viscosity coefficient, for regions of the flow with strong adverse pressure gradients \citep{Matyushenko2016}. To illustrate the effect of their proposed change, we present \gls{rans} simulations with the baseline and proposed values of $a_1$. Figure \ref{fig:adhoc_a1} indicates that decreasing $a_1$ from its baseline value of 0.31 to the value of 0.28 as recommended by \cite{Matyushenko2016} leads to the prediction of a separation (albeit too strong), as indicated by the flattening of the coefficient of pressure $C_p$ in the vicinity of $x/L=0.2$. However, the skin friction coefficient $C_f=\tau_w/(\rho_\infty U_\infty^2/2)$ is underpredicted in the upstream and downstream regions of the flow, where $\tau_w$ is the wall shear stress. This indicates that the proposed change degrades the calibration of the baseline model for the \gls{zpg} \gls{bl}, which is consistent with the findings of \cite{Matyushenko2016}.

\begin{figure}
\centering
\floatsetup[figure]{style=plaintop} 
\captionsetup[subfigure]{position=top, justification=raggedright, singlelinecheck=false} 
\subcaptionbox{}{\includegraphics[width=0.49\textwidth]{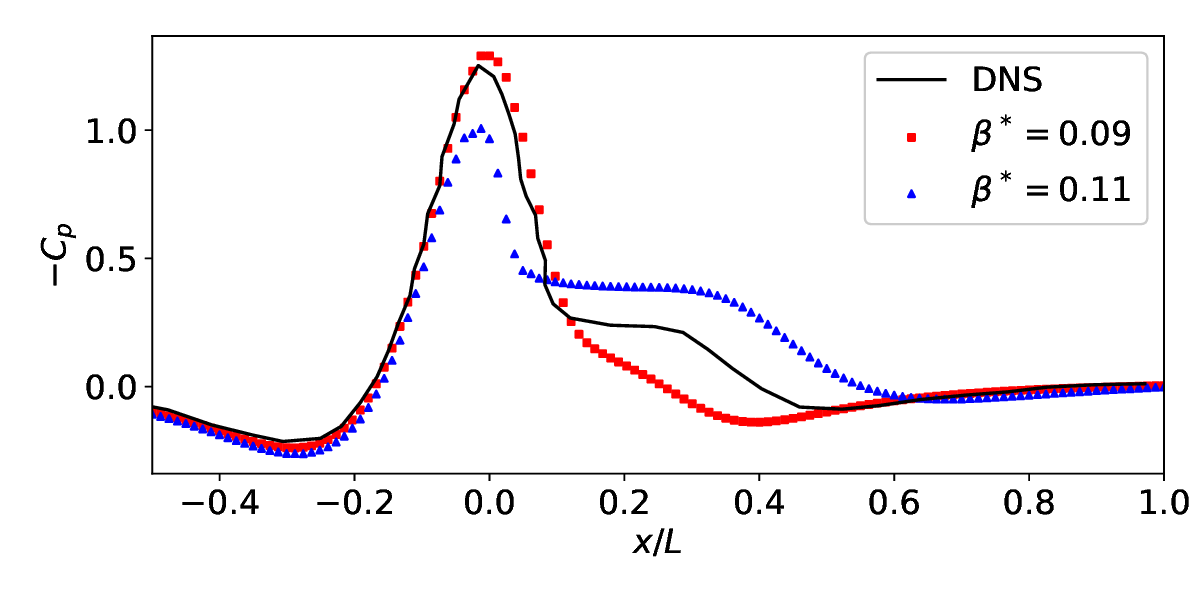}}
\subcaptionbox{}{\includegraphics[width=0.49\textwidth]{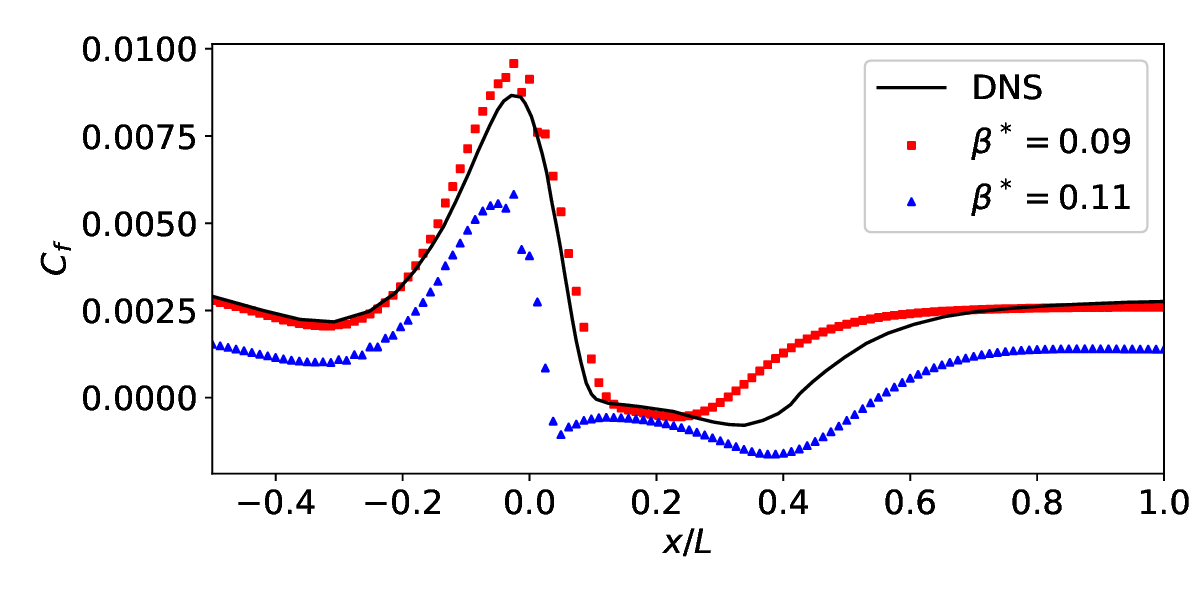}}
\caption{Effect of ad hoc tuning of $\beta^*$ in the \gls{sst} model is shown for the Boeing Gaussian bump by results with the baseline value $\beta^*=0.09$ in red and with the elevated value suggested by \cite{Zhong2018} $\beta^*=0.11$ shown in blue. Reference \gls{dns} data are shown in black.}
\label{fig:adhoc_betas}
\end{figure}

A second variant uses a modified value of $\beta^*$ in the $k$ equation \citep{Zhong2018}. This coefficient appears in the $k$-destruction term, the boundary layer sensors through $arg_1$ and $arg_2$, and in the production limiter. As shown in Fig.~\ref{fig:adhoc_betas}, increasing $\beta^*$ from its baseline value of 0.09 to the value of 0.11 as recommended by \cite{Zhong2018} leads to similar effects as modifying $a_1$; namely, the separation is predicted (albeit too large), but a significant error in $C_f$ is introduced in the nearly \gls{zpg} regions upstream and downstream of the bump.

A third variation of the \gls{sst} model is the generalized $k$-$\omega$ (GeKO), which is a proprietary model developed by ANSYS. For three terms in the $\omega$ transport equation, GeKO introduces multiplicative functions parameterized by six user-specified coefficients \citep{Menter2019}. The user is encouraged to tune the coefficients to achieve desired performance, e.g., the prediction of separation in some benchmark test problem. The multiplicative functions are designed to be invariant to the user-specified coefficients in \gls{zpg}\glspl{bl} to recover the calibration of the baseline model in this case. However, case-by-case tuning of coefficients implies a nonuniversal model, which limits the confidence of applying the model to novel geometries \cite{Spalart2023}.

The shortcomings of previous models can be summarized as violating the following list of model design objectives:
\begin{enumerate}
    \item Model should improve the prediction of separation
    \item Model should accurately predict \gls{zpg}\glspl{bl}
    \item Model coefficients should not change with modest changes in Reynolds number (e.g., a factor of 10) or modest changes in geometry (e.g., across two different airfoil shapes).
\end{enumerate}

The remainder of this paper is dedicated to developing two models that satisfy the modeling objectives outlined above. The paper is organized as follows: The computational setups of \textit{a posteriori} \cgls{rans} calculations of the flows over the Boeing Gaussian bump, the FFA-W3-301 airfoil, and the DU00-W-212 airfoil are discussed in Section \ref{sec:setup}; \cgls{dns} data for the Boeing Gaussian bump are analyzed in Section \ref{sec:model} to guide the physics-driven design of the pressure gradient sensor and the definition of two proposed \cgls{rans} models; the data-informed calibration of the model coefficients and application of the models to the Boeing Gaussian bump, the FFA-W3-301 airfoil, and the DU00-W-212 airfoil are presented in Section \ref{sec:results}; and conclusions are offered in Section \ref{sec:conclusion}.

\section{Computational methodology} \label{sec:setup}
\gls{rans} calculations are performed using Nalu-Wind, an incompressible solver, which is part of the ExaWind open-source high-fidelity software suite developed by \gls{nrel} and Sandia National Laboratories. ExaWind has been developed for modern high-performance computing systems, is compatible with graphical-processing units (GPUs), and has been extensively validated for wind energy applications \citep{Sprague2020, Sharma2024}. Nalu-Wind is a 3D unstructured unsteady second-order-accurate finite volume solver. Systems of linear equations are solved using the generalized minimal residual method with the BoomerAMG algebraic multi-grid preconditioner via {\it hypre}, a software library of high-performance preconditioners and solvers. The no-slip walls are treated with turbulence boundary conditions following \cite{Menter1993}. Far-field turbulence is specified with the turbulence boundary conditions $k_{ff}=10^{-3} U_\infty^2/Re_L$ and $\omega_{ff}=5 U_\infty /L$ in line with the recommendations of \cite{Menter1993}. $Re_L=\rho L U_\infty/\mu$ is the Reynolds number based on the characteristic length scale of the geometry $L$ and the freestream velocity $U_\infty$.

\subsection{Computational setup for airfoil calculations}
We consider the flow over the FFA-W3-301 airfoil at $Re_c=1.6$ million and the flow over the DU00-W-212 airfoil at $Re_c=(3,6,9,12,$ and $15)\times10^6$. These airfoils are chosen because they have maximum thicknesses of 30\% and 20\% of their chord lengths, respectively, which are representative of the cross sections of very large (greater than 10 MW) wind turbines. Large turbines use thick airfoil cross sections to reduce rotor weight \citep{Grasso2014}. Beyond their applicability, thick airfoils are particularly relevant for this study because the baseline SST model has been shown to poorly predict stall, especially on thick airfoils \citep{Gutierrez2023}. 

\begin{figure}
\centering
\floatsetup[figure]{style=plaintop} 
\captionsetup[subfigure]{position=top, justification=raggedright, singlelinecheck=false} 
\subcaptionbox{}{\includegraphics[width=0.49\textwidth]{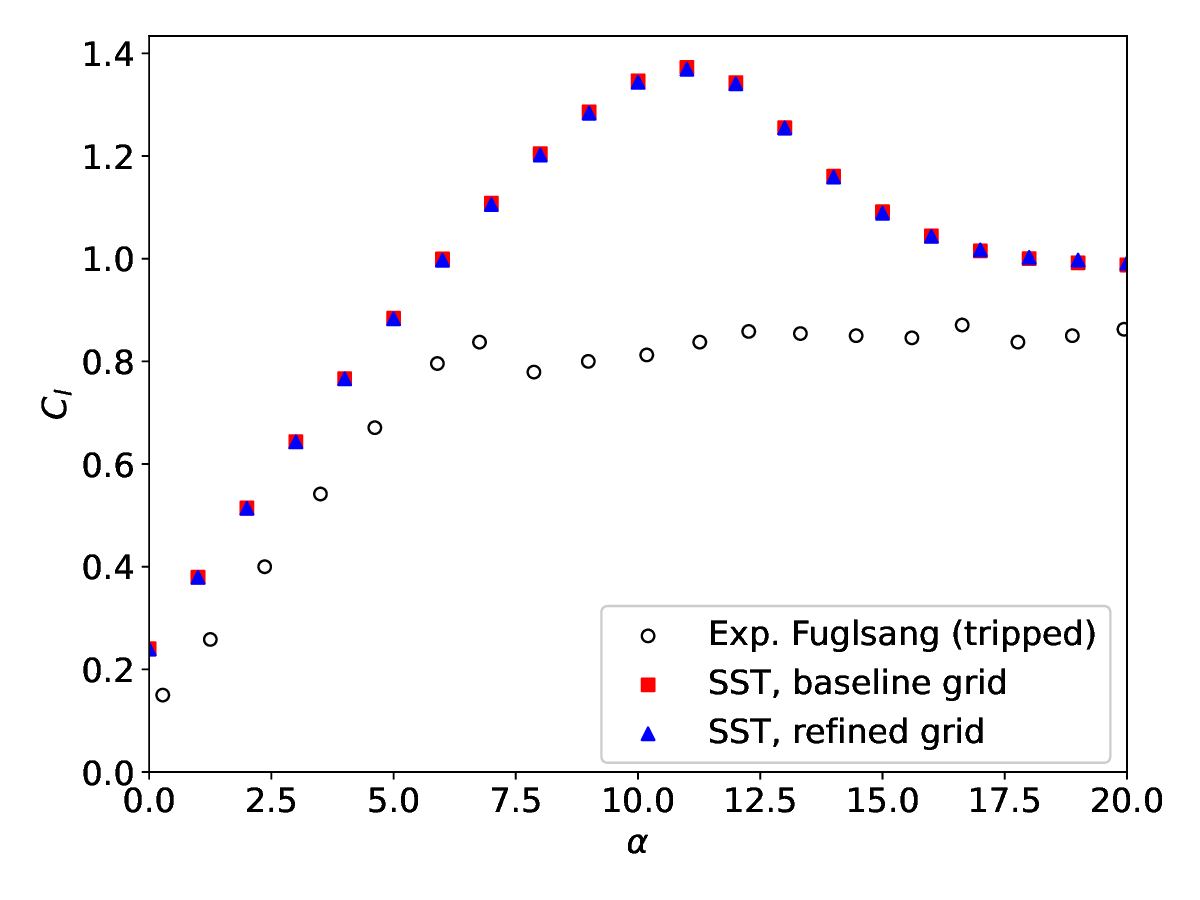}}
\subcaptionbox{}{\includegraphics[width=0.49\textwidth]{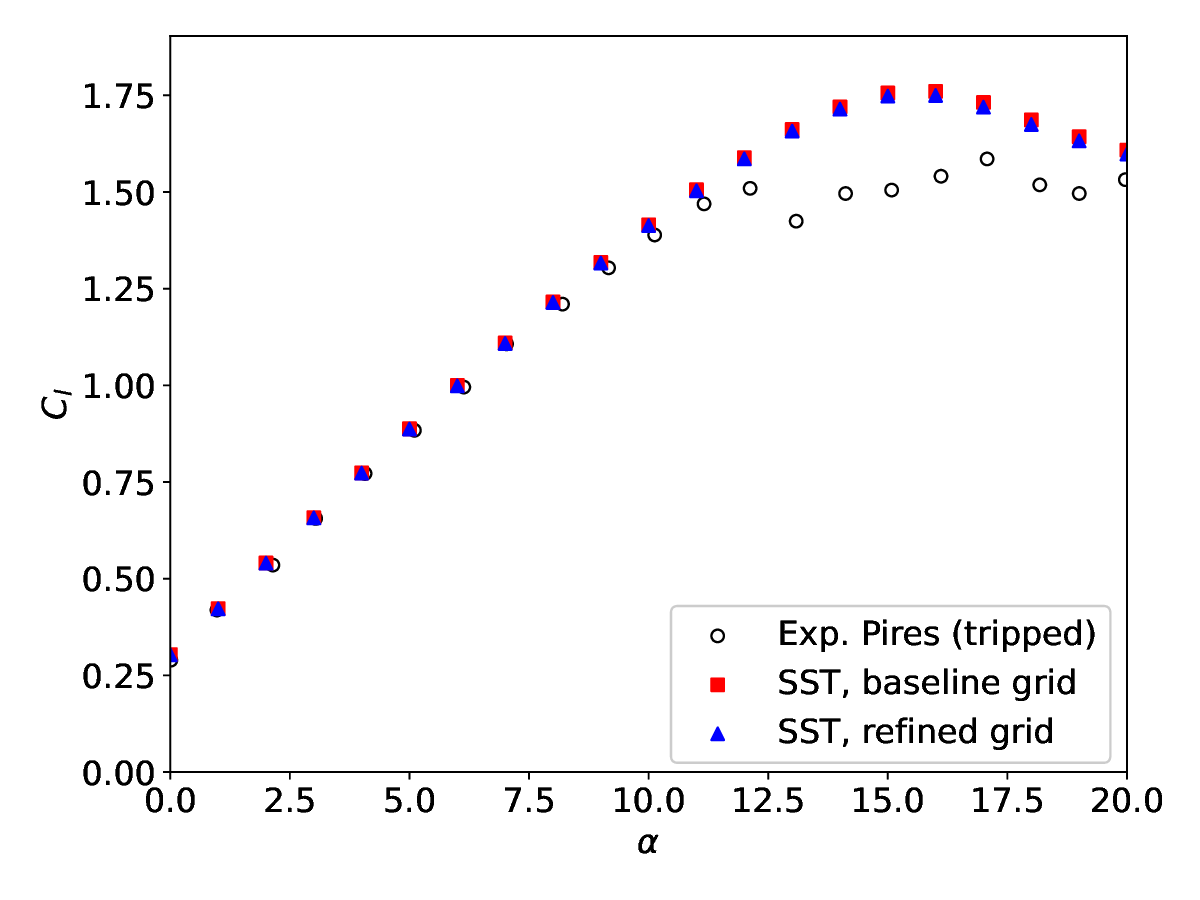}}
\caption{Mesh convergence of \gls{rans} simulations of the FFA-W3-301 airfoil (a) at $Re_c=1.6\times 10^6$ and the DU00-W-212 airfoil (b) at $Re_c=1.5\times 10^7$ compared to reference experiments \citep{Fuglsang1998,Pires2016} (black). Coefficient of lift is plotted versus the angle of attack for the baseline grid (in red) and for a spatially refined grid (in blue).}
\label{fig:airfoil_conv}
\end{figure}

2D steady RANS calculations are employed. The computational grids are O-grids with 385 points each on the suction and pressure sides following a hyperbolic tangent distribution with minimum spacing of $\Delta_t/L=6.1\times10^{-4}$ at the leading edge and $\Delta_t/L=3.3\times10^{-4}$ at the trailing edge. There are 162 geometrically spaced wall-normal points with a minimum wall-normal spacing of $\Delta_y/L=2.1\times 10^{-7}$ and a stretching ratio of 1.13. 
For the FFA-W3-301 airfoil, and there are a total of 133,000 grid cells. For simulations with the baseline SST model with $0^\circ \le \alpha \le 20^\circ$, the maximum wall-normal grid spacing at the wall $\Delta_y^+<0.022$. 
For the DU00-W-212 airfoil, there are a total of 127,000 grid cells. For simulations with the baseline SST model with $0^\circ \le \alpha \le 20^\circ$, the maximum wall-normal grid spacing at the wall $\Delta_y^+<0.037,0.069,0.10,0.13,0.16$ for flows with $Re_c=3,6,9,12,15\times10^6$, respectively.
In Fig.~\ref{fig:airfoil_conv}, the coefficient of lift is plotted for the baseline grids described above and for grids that are refined at the airfoil surface by a factor of two in the wall-normal direction. Agreement between resolutions suggests the results are spatially converged; the discrepancy of the converged simulations with respect to the experimental data motivates the model development in this paper.

\subsection{Computational setup for the Boeing Gaussian bump}
The Boeing Gaussian bump has been widely studied in the literature, including experimental investigations \citep{Williams2020,Williams2021,Gray2021,Robbins2021}, \glspl{dns} \citep{Balin2020,Uzun2021,Uzun2022,Prakash2024}, large-eddy simulations \citep{Balin2020a,Iyer2021,Prakash2022,Whitmore2021,Zhou2023}, and \gls{rans} simulations \citep{Williams2020,Balin2020a,Iyer2021,Prakash2022,Prakash2024}. This flow was designed in collaboration with Boeing to represent smooth-body aircraft separation \citep{Williams2020} and poses a significant challenge to existing \gls{rans} technology. Indeed, to the authors' knowledge, the accurate capture of separation in this flow using \gls{rans} simulations has not been reported in the open literature to date, which motivates the study of this flow.

\begin{figure}
\centering
\includegraphics[width=0.49\textwidth]{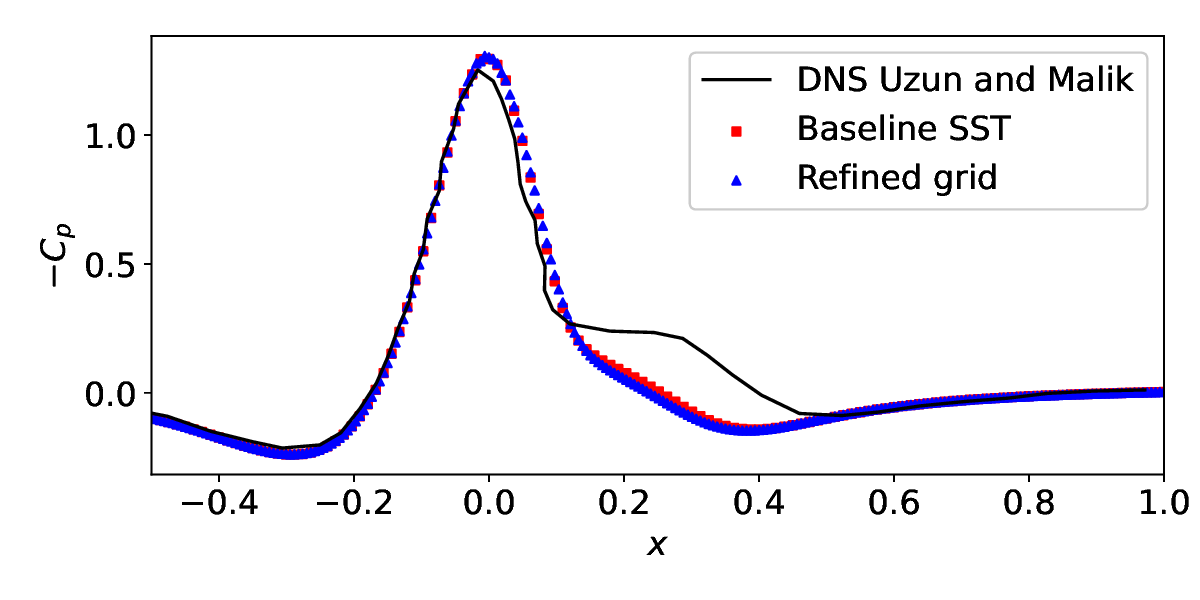}
\caption{Mesh convergence of \gls{rans} simulations of the Boeing Gaussian bump compared to DNS data \citep{Uzun2022} (black). Coefficient of pressure is plotted versus the streamwise coordinate for the baseline grid (in red) and for a spatially refined grid (in blue). For reference, DNS data \citep{Uzun2022} are shown in black.}
\label{fig:bump_conv}
\end{figure}

Comparison is made to the spanwise-periodic \gls{dns} investigation of \cite{Uzun2022} with a freestream Reynolds number of 2 million based on the characteristic length scale of the bump \citep{Williams2020}. 
The boundary conditions of the present \gls{rans} simulations are chosen to best match the \gls{dns} as follows: 
The geometry of the bump surface is given by $y/L=0.085 \exp(-(x/(0.195L))^2)$; thus, $x=0$ is the apex of the bump.
Following the recommendations of the NASA Turbulence Modeling Resource~\citep{Rumsey2023}, the inflow is a uniform flow over a symmetry (slip) wall of streamwise length $L=0.5$. This extends from $-1.556<x/L<-1.056$.
For $x/L>-1.056$, the no-slip condition is applied. In setting up this case, the length of this no-slip wall upstream of the bump apex was varied iteratively, and it was found that this choice led to the development of a boundary layer thickness downstream at $x/L=-0.65$ that agrees well with the DNS data. Specifically, the boundary layer thickness is computed using the method of \cite{Griffin2021} to be $\delta_{99} = 0.0072$ for the \gls{dns} and the $\delta_{99} = 0.0072$ for the \gls{rans}.
The top boundary is modeled with a symmetry boundary condition for velocity at $y=L$ with the far-field turbulence values that were specified above.
The computational grid consists of 288 uniformly spaced points in the streamwise direction and 222 geometrically stretched points in the wall-normal direction with a stretching ratio of 1.04 following \cite{Bidadi2023} for a total of 63,427 grid cells. The streamwise grid spacing $\Delta_x=0.0125L$ and the minimum wall-normal spacing is $\Delta_y=6.875\times 10^{-6}L$, which is less than unity in viscous units for all wall-adjacent grid points.
In Fig.~\ref{fig:bump_conv}, the coefficient of pressure is plotted for the baseline grid described above and for a grid that is refined at the bump surface by a factor of two in both the streamwise and wall-normal directions. Agreement between resolutions suggests the results are spatially converged; the discrepancy of the converged simulations with respect to the DNS data motivates the model development in this paper.

To verify our setup of this case, the \gls{sst} results of \cite{Prakash2024} are compared against our results using Nalu-Wind and the baseline \gls{sst} model in Fig.~\ref{fig:bump_baseline}. As expected, the predictions of $C_p$ and $C_f$ by the \gls{sst} model agree fairly well, but neither captures the separation observed in the \gls{dns} data of \cite{Uzun2022}.

\section{Model formulations} \label{sec:model}
Due to its extensive development and widespread implementation and utilization, the 2003 version of the Menter \gls{sst} model \citep{Menter2003} is the starting point for the proposed model. We consider the work of \cite{Matyushenko2016} and \cite{Zhong2018}, which proposed modifying $a_1$ and $\beta^*$, respectively. Both of these approaches satisfy our first objective of improving the prediction of separation but also violate our second objective of retaining accuracy in \gls{zpg}\glspl{bl} as demonstrated in Figs. \ref{fig:adhoc_a1} and \ref{fig:adhoc_betas}. In addition, there is some evidence that the $a_1$ approach violates the third objective insofar as the choice of $a_1$ that leads to the best prediction of separation varies between airfoil geometries \citep{Gutierrez2023}. Thus, we propose the introduction of an adverse pressure gradient sensor, which can confine any modified coefficients to regions of strong adverse pressure gradients, thereby leaving the baseline \gls{sst} model unchanged in regions of \gls{zpg} or favorable pressure gradients, where the \gls{sst} model is already well calibrated. 


\subsection{Developing a pressure gradient sensor}
It should be noted that the baseline 2003 version of the \gls{sst} model \citep{Menter2003} nominally already has an APG sensor given by
\begin{equation}
    I_{03} \equiv \left(a_1 \omega < S F_2 \right),
\end{equation}
which was updated from the original 1993 version of the sensor \citep{Menter1993}
\begin{equation}
    I_{93} \equiv \left(a_1 \omega < \Omega F_2 \right),
\end{equation}
where the vorticity magnitude $\Omega = \sqrt{2 W_{ij} W_{ij}}$ and the rate of rotation tensor $W_{ij} = [\pderi{u_i}{x_j}-\pderi{u_j}{x_i}]/2$.

Both of these sensors will be shown to exhibit shortcomings in robustly identifying adverse pressure gradients. This motivates the development of the following pressure gradient sensor
\begin{equation} \label{proposed:sst_apg}
    I_{p} \equiv \left(\widehat{u_i-U_{i,w}} \pder{p}{x_i} F_2 > s_T \frac{\rho_\infty U_\infty^2}{L}\right),
\end{equation}
where $L$ is the characteristic length scale of the geometry (e.g., the chord length of an airfoil or the characteristic length scale of the Boeing Gaussian bump \citep{Williams2020}) and $s_T$ is a model constant that determines the strength of the nondimensional pressure gradient that is sufficiently adverse to be labeled as an APG region. Recall that $F_2$ tends to unity inside boundary layers. $U_{i,w}$ is the velocity of the nearest wall (viscous solid surface) to enforce Galilean invariance. For the remainder of this work, $U_{i,w}=0$. The unit vector operator is denoted by~$\widehat{\cdot}$, i.e., $\widehat{\phi}_i = \phi_i/(\sqrt{\phi_j \phi_j} + \varepsilon)$, where $\varepsilon$ is a small value added for numerical robustness at stagnation, separation, and reattachment points. Further study could be devoted to other nondimensionalizations of the pressure gradient when $L$ and $U_\infty$ are less clearly known, but for the simple geometries considered in this work, this nondimensionalization is simple and well established. The velocity unit vector is used so as to extract the pressure derivative in the direction of the local mean flow.

\begin{figure}
\centering
\floatsetup[figure]{style=plaintop} 
\captionsetup[subfigure]{position=top, justification=raggedright, singlelinecheck=false} 
\subcaptionbox{}{\includegraphics[width=0.49\textwidth]{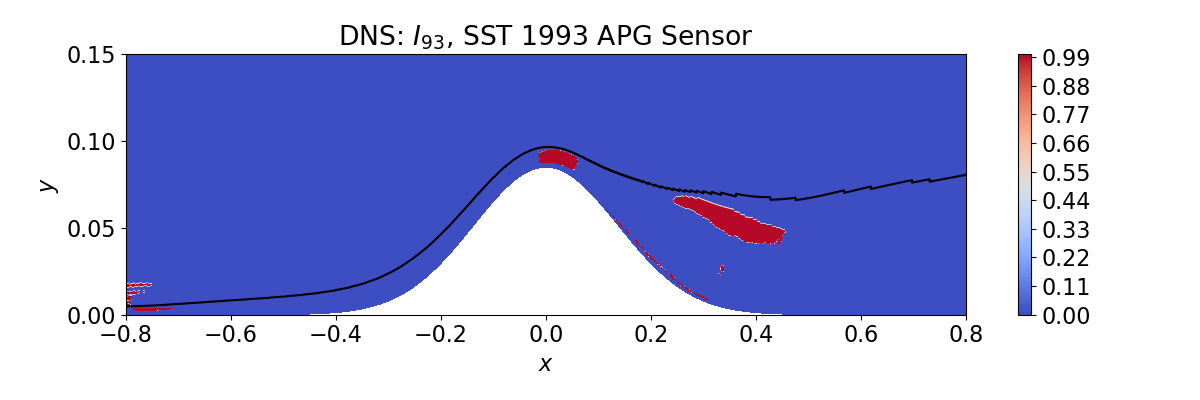}}
\subcaptionbox{}{\includegraphics[width=0.49\textwidth]{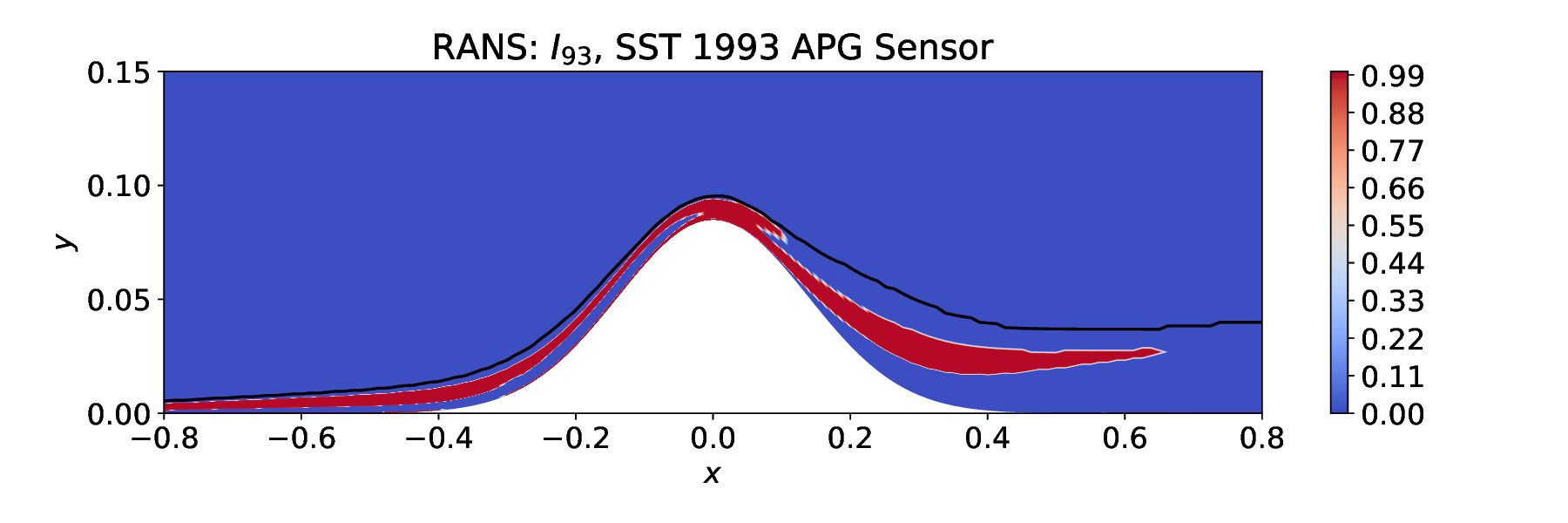}} \\
\subcaptionbox{}{\includegraphics[width=0.49\textwidth]{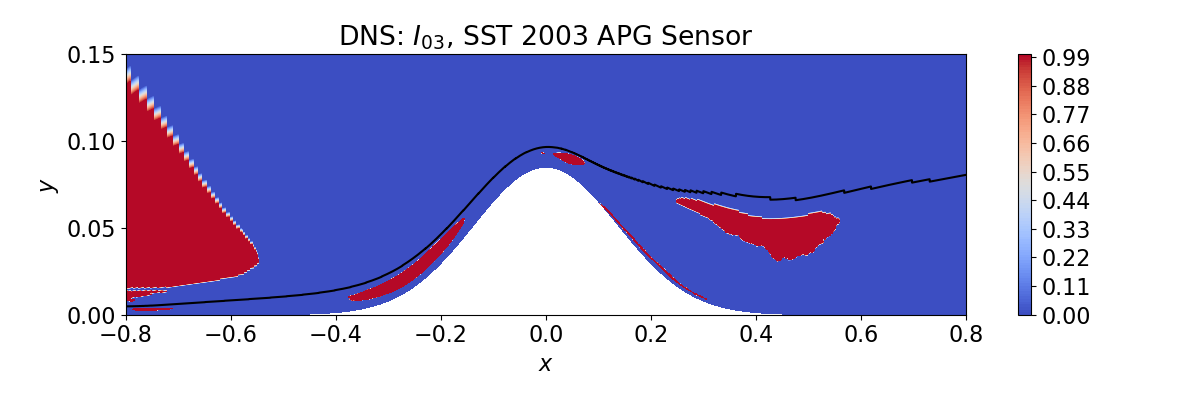}}
\subcaptionbox{}{\includegraphics[width=0.49\textwidth]{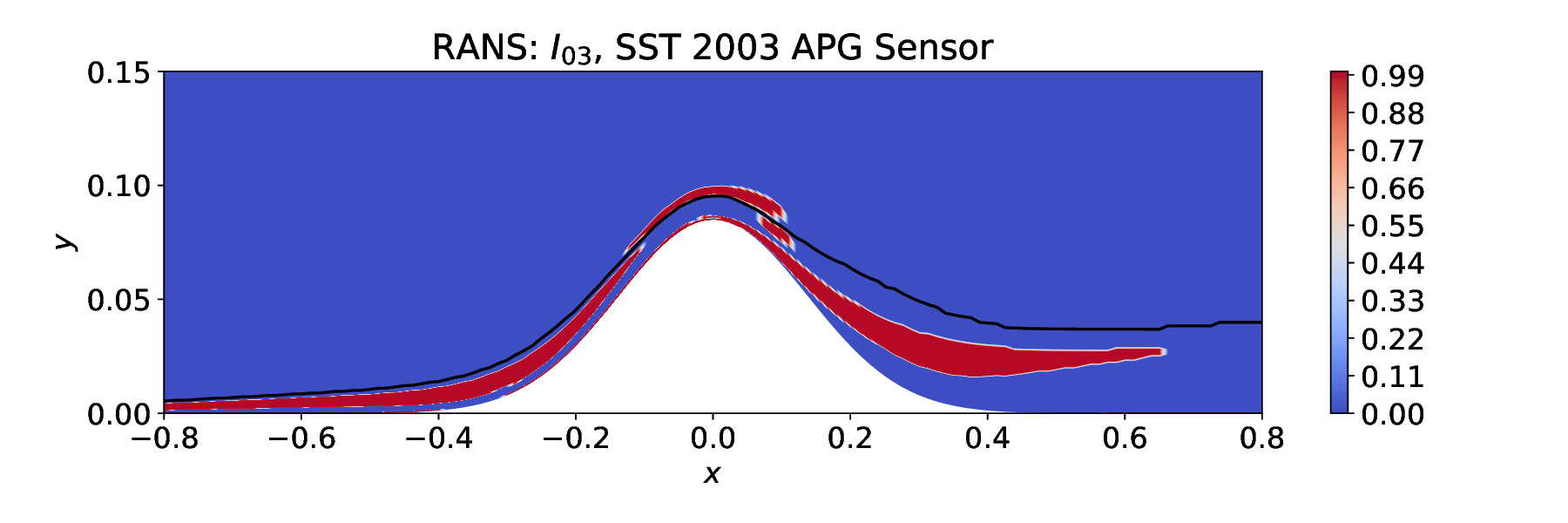}} \\
\subcaptionbox{}{\includegraphics[width=0.49\textwidth]{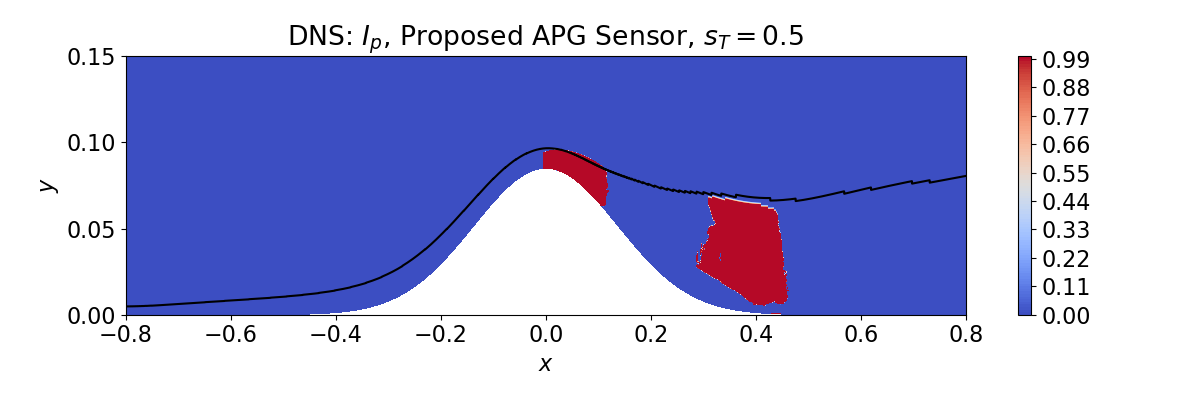}}
\subcaptionbox{}{\includegraphics[width=0.49\textwidth]{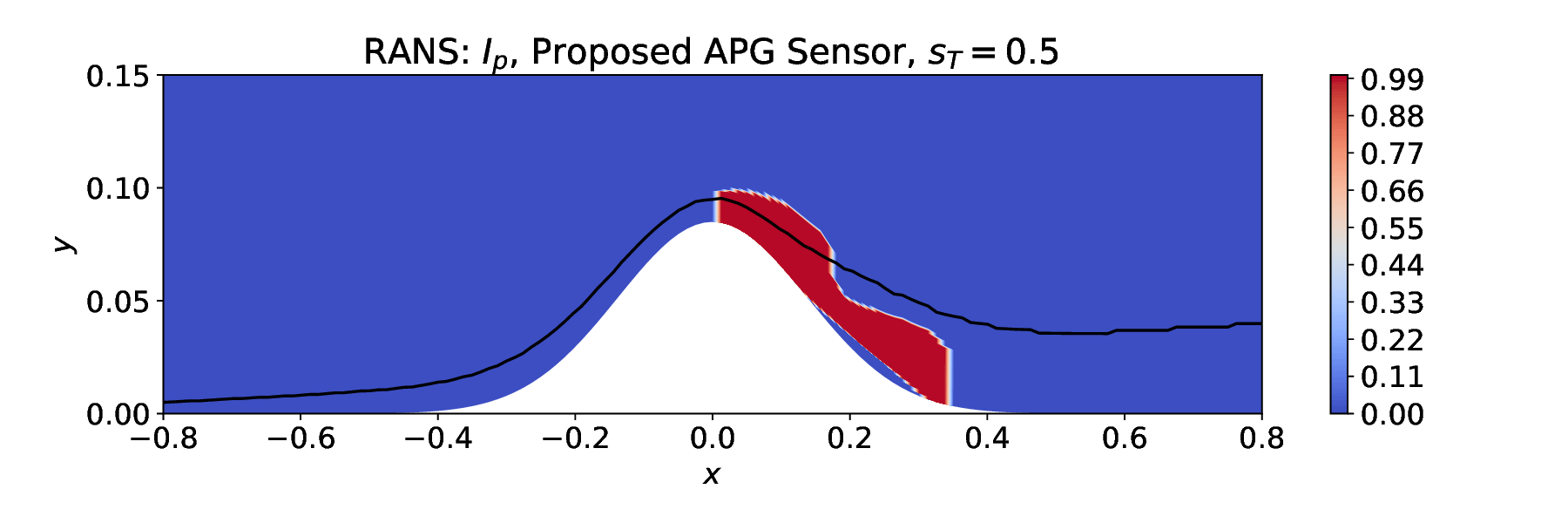}}
\caption{Pressure gradient sensors of \cite{Menter1993} (a,b), \cite{Menter2003} (c,d), and the present work (e,f) on the Boeing Gaussian bump. Evaluations of these sensors on DNS data \citep{Uzun2022} and RANS predictions using the baseline SST model with Nalu-Wind are presented in the left and right columns, respectively. Red indicates a region identified as strong APG; blue indicates otherwise. The black line indicates the boundary layer thickness \citep{Griffin2021}.}
\label{fig:apg}
\end{figure}

Considering the flow over the Boeing Gaussian bump, the proposed and existing sensors are plotted for both \gls{dns} and \gls{rans} data in Fig.~\ref{fig:apg}, where the sensor value of unity is represented by the red region and corresponds to a region flagged as \gls{apg}. Note that the same RANS simulation using the Menter 2003 turbulence model is used to generate Fig.~\ref{fig:apg}b, d, and f; thus, the figures are only exposing differences in the sensor definitions, since the underlying flows are identical. 
Figure~\ref{fig:apg}a, c, and e provide estimates of the RANS sensors using DNS data; we begin by computing $k$ and $\mu_t$ from the mean velocities and Reynolds stresses. In order to estimate $\omega$ from DNS data, we rearrange Eq.~\ref{eq:mut}. However, this would lead to regions of space where $\omega$ is not defined. To extend $\omega$ to be defined everywhere (as it is in RANS), we compute it as $\omega = \rho k/\mu_t$ at all points in space. 
The slope of the wall-pressure \gls{dns} data in Fig.~\ref{fig:bump_baseline}a clearly indicates a strong favorable pressure gradient for $-0.25<x<0$.
However, in Fig.~\ref{fig:apg}c and d, the SST 2003 sensor $I_{03}$ labels this region as \gls{apg} for both the \gls{dns} and \gls{rans} data. Although the Menter 1993 sensor $I_{93}$ correctly identifies this region in the \gls{dns} data (Fig.~\ref{fig:apg}a), the \gls{rans} data incorrectly labels it as \gls{apg} (Fig.~\ref{fig:apg}b).
The proposed sensor, unlike the existing \gls{sst} sensors, identifies only the region of strong \gls{apg} consistent with the slope of the wall pressure data in Fig.~\ref{fig:bump_baseline}a. In Fig.~\ref{fig:apg}e, the sensor identifies the portions of the separation bubble near separation and near reattachment. In Fig.~\ref{fig:apg}f, the entire separation bubble is identified. Note that the underlying SST 2003 RANS data for Fig.~\ref{fig:apg}b, d, and f predict reattachment about $0.1L$ before the DNS (see Fig.~\ref{fig:bump_baseline}).

\begin{figure}
\centering
\floatsetup[figure]{style=plaintop} 
\captionsetup[subfigure]{position=top, justification=raggedright, singlelinecheck=false} 
\subcaptionbox{}{\includegraphics[width=0.49\textwidth]{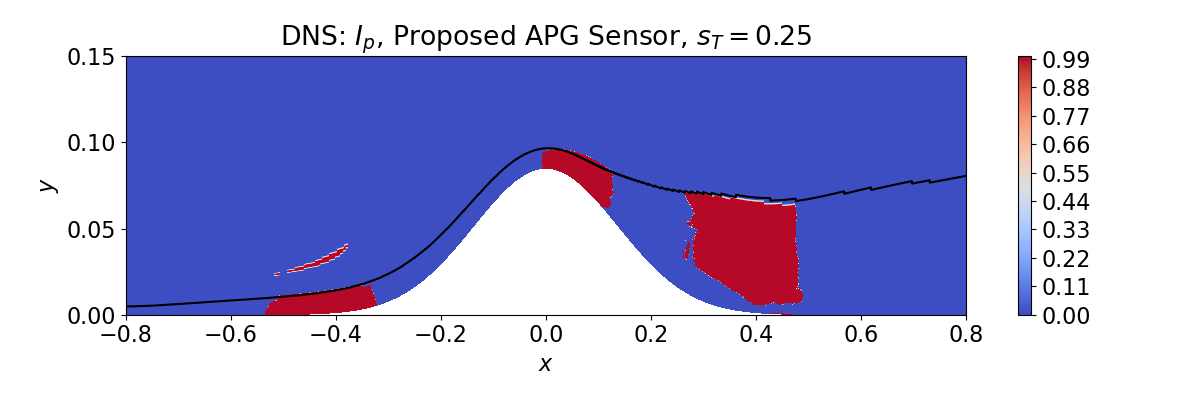}}
\subcaptionbox{}{\includegraphics[width=0.49\textwidth]{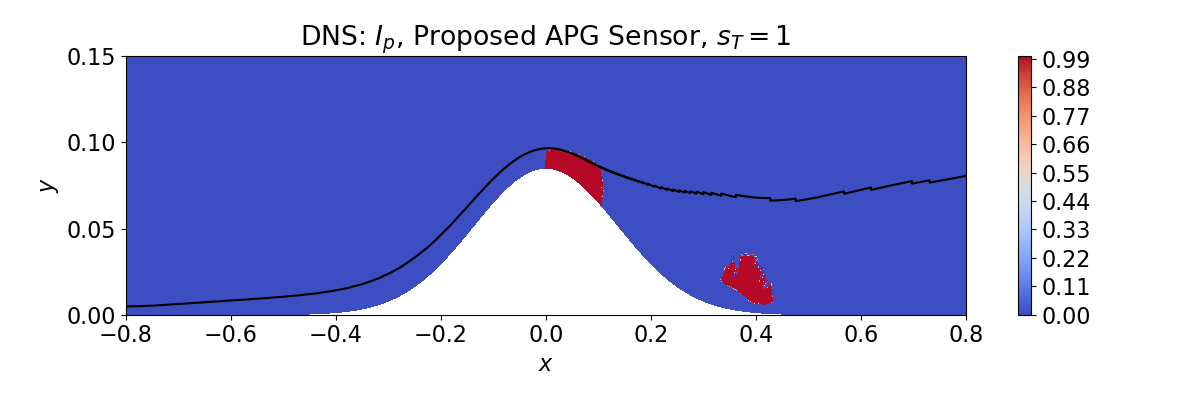}} \\
\subcaptionbox{}{\includegraphics[width=0.49\textwidth]{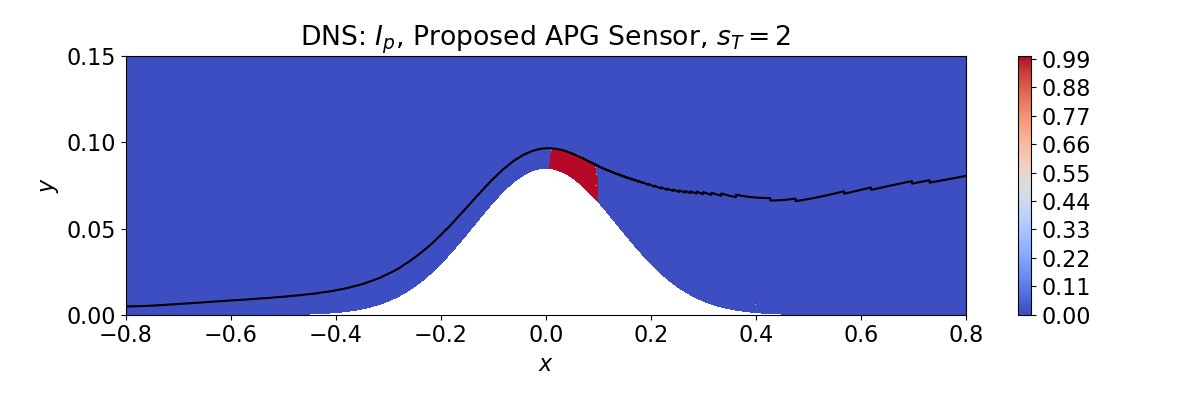}}
\subcaptionbox{}{\includegraphics[width=0.49\textwidth]{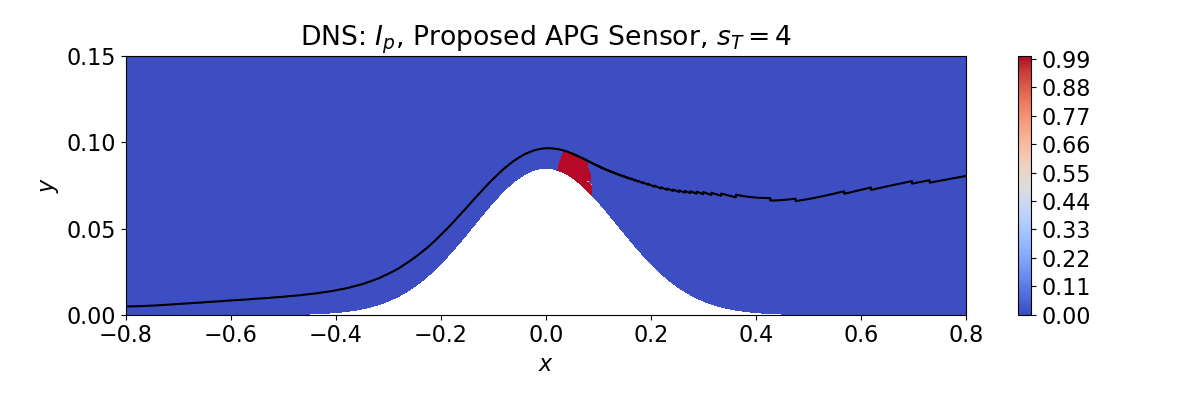}}
\caption{The effect of varying the threshold constant $s_T$ for contours of the proposed \gls{apg} sensor evaluated with DNS data \citep{Uzun2022} for the Boeing Gaussian bump. Results for $s_T=0.25$, $1$, $2$, and $4$ are shown in panels a, b, c, and d, respectively. Red indicates a region identified as strong APG; blue indicates otherwise. The black line indicates the boundary layer thickness \citep{Griffin2021}.}
\label{fig:sT}
\end{figure}

The proposed sensor uses a threshold constant $s_T$, which determines the strength at which the \gls{apg} is considered strong and the turbulence model will need to be augmented. The effect of varying this constant is explored in Fig.~\ref{fig:sT} (compare also to Fig.~\ref{fig:apg}e). Here, we consider DNS data to avoid muddling our analysis with errors in the baseline SST model's solution. In Fig.~\ref{fig:sT}a, $s_T=0.25$ is considered, and the foot of the bump near $x=-0.4$ is identified as a strong \gls{apg}. Meanwhile, Fig.~\ref{fig:bump_baseline}a indicates that the \gls{dns} wall pressure data have only a mildly adverse slope in this region. Moreover, in this mild pressure gradient, the baseline \gls{sst} model provides a good prediction of $C_p$ and $C_f$, as shown in Fig.~\ref{fig:bump_baseline}. Our second modeling objective indicates that $s_T$ should be larger to preserve the model's favorable performance in this region. On the other hand, for $s_T=0.5, 1, 2,$ and $4$ (Fig.~\ref{fig:apg}e and Fig.~\ref{fig:sT}b, c, d) the identified regions are only places where the baseline SST is failing to accurately predict $C_p$ and/or $C_f$ (see Fig.~\ref{fig:bump_baseline}). 
Higher values of $s_T$ will lead to a more targeted augmentation of the \gls{sst} model. It is expected that if the proposed eddy viscosity models promote separation, then lower values of $s_T$ will apply these models over more of the flow and lead to earlier separation (e.g., stall will occur at a lower angle of attack). Since Fig.~\ref{fig:bump_baseline} indicates that the baseline SST model is performing well for the upstream section of the bump ($x\lessapprox0$), it is likely that the APG sensor should not label this region as a strong APG region needing model augmentation; thus, we expect to find that $s_T\ge0.5$ will lead to the best results for the Boeing Gaussian bump. The exact value will be determined via \textit{a posteriori} calibration in Section \ref{sec:results}.

\subsection{Developing two modified eddy viscosity models for APG flow}
To develop variants of the \gls{sst} model that might satisfy our three modeling objectives from Section \ref{sec:intro}, we consider the proposed methods of \cite{Matyushenko2016} and \cite{Zhong2018} of varying the $a_1$ and $\beta^*$ \gls{sst} coefficients, respectively. We propose to only modify the coefficients in regions indicated by our pressure gradient sensor, i.e., $I_p=\mathrm{True}$, and use the default coefficient values otherwise. This yields the following two proposed models:
The $a_{1,\mathrm{APG}}$ model is the same as the baseline SST, except that Eq.~\ref{eq:mut} is replaced with the following:
\begin{equation}
\mu_t =
\begin{cases}
  \frac{\rho k}{\omega} & \text{if } a_1 \omega > S F_2, \\
  \frac{a_{1}' \rho k}{S F_2} & \text{otherwise},
\end{cases}
\label{eq:mut_mod}
\end{equation}
where
\begin{equation}
a_{1}' =
\begin{cases}
  a_{1,\mathrm{APG}} & \text{if } \left(\widehat{u}_i \pder{p}{x_i} F_2 > s_T \frac{\rho_\infty U_\infty^2}{L}\right), \\
  a_1 & \text{otherwise}.
\end{cases}
\label{eq:a1_model}
\end{equation}
The $\beta^*_\mathrm{APG}$ model is the same as the baseline SST, except that Eq.~\ref{eq:k} is replaced with the following:
\begin{equation} \label{eq:k_mod}
    \pder{\rho k}{t} + \pder{\rho u_j k}{x_j} = P - \beta^{*\prime} \rho \omega k + \pder{}{x_j}[(\mu + \sigma_k \mu_t) \pder{k}{x_j}],
\end{equation}
where
\begin{equation}
\beta^{*\prime} =
\begin{cases}
  \beta^*_\mathrm{APG} & \text{if } \left(\widehat{u}_i \pder{p}{x_i} F_2 > s_T \frac{\rho_\infty U_\infty^2}{L}\right), \\
  \beta^* & \text{otherwise},
\end{cases}
\label{eq:betas_model}
\end{equation}
where $a_{1,\mathrm{APG}}$ and $\beta^*_\mathrm{APG}$ are augmented model coefficient values that should promote separation. 
The baseline model and the two proposed models are summarized in Table \ref{tab:models}.
\begin{table}[]
\begin{tabular}{l|llll}
\textbf{Model}               & \textbf{$a_1^\prime$ definition} & \textbf{$\beta^{*\prime}$ definition} &  &  \\ \cline{1-3}
Baseline model               & $a_1^\prime=a_1$                 & $\beta^{*\prime}= \beta^{*}$          &  &  \\ \cline{1-3}
$a_{1,\mathrm{APG}}$ model   & Eq.~\ref{eq:a1_model}            & $\beta^{*\prime} = \beta^{*}$         &  &  \\ \cline{1-3}
$\beta^*_\mathrm{APG}$ model & $a_1^\prime=a_1$                 & Eq.~\ref{eq:betas_model}              &  & 
\end{tabular}
\caption{Summary of baseline and proposed model definitions.} \label{tab:models}
\end{table}

Recall that $F_2$ tends to unity inside boundary layers. Less eddy viscosity will make the velocity profile less full and more susceptible to separation. When $a_1 \omega < SF_2$, $\tau_w \sim \mu_t S \sim \rho a_1 k$, so reducing $a_1$ or reducing $k$ by increasing $\beta^*$ (the $k$-destruction coefficient) are model augmentations that will likely promote separation in line with our first modeling objective. By the construction and analysis of the pressure gradient sensor, the model satisfies our second objective that it should revert to the baseline \gls{sst} model in non-\gls{apg} flows, assuming $s_T$ is calibrated appropriately. 
The conditional in the first branch of Eq.~\ref{eq:mut_mod} is still written in terms of $a_1$ (rather than $a_{1}'$) since this is found to be a key component of the baseline SST model's calibration, which we do not want to compromise.
$\beta^*$ appears in many places in the SST model, but we only replace $\beta^*$ with $\beta^{*\prime}$ as the coefficient for the $k$-destruction term; $arg_1$, $arg_2$, and the production limiter remain defined in terms of $\beta^*=0.09$ (i.e., $\beta^{*\prime}$ is not used). It was found that these terms do not play a dominant role in the onset of separation, and the model's action could be restricted to the $k$-destruction term.

$I_{93}$ and $I_{03}$ use $a_1$ both as the sensor threshold constant (similar to $s_T$) and as the turbulence model constant since it contributes to the value of $\mu_t$ in APG regions. Meanwhile, the proposed models break these distinct functions into a pair of coefficients ($s_T$ and $a_{1,\mathrm{APG}}$ or $s_T$ and $\beta^*_\mathrm{APG}$). This allows for independent, and thus more precise, calibration of the independent roles of the APG threshold and the turbulence transport equations in APG regions.

\section{Results} \label{sec:results}
In this section, the coefficients $s_T$, $a_{1,\mathrm{APG}}$, and $\beta^*_\mathrm{APG}$ will be specified for our two proposed models in Eqs.~\ref{eq:a1_model} and \ref{eq:betas_model}. This constitutes the data-informed part of the modeling approach, as the coefficients are selected so as to obtain the best agreement of \textit{a posteriori} RANS results and experimental data for the FFA-W3-301 airfoil. The effect of the models on the prediction of separation will be assessed through their application to the simulation of the FFA-W3-301 and DU00-W-212 airfoils and the Boeing Gaussian bump.

\subsection{Airfoil results}
Both models use the pressure gradient sensor, which introduces the undetermined constant $s_T$, which controls the region where the adverse pressure gradient is deemed strong enough to merit the activation of the proposed eddy viscosity models. We perform a grid search by running simulations with a range of reasonable choices of $s_T$ and $a_{1,\mathrm{APG}}$ for the $a_{1,\mathrm{APG}}$ model and $s_T$ and $\beta^*_\mathrm{APG}$ for the $\beta^*_\mathrm{APG}$ model. The search bounds are informed by our \textit{a priori} investigation of the effect of $s_T$ on the Boeing Gaussian bump and by the investigations of $a_1$ and $\beta^*$ in \cite{Matyushenko2016} and \cite{Zhong2018}. Since the simulations run in minutes on a single core and cases can be run synchronously on a multi-core node, the search is not computationally demanding. The error metric is the lift coefficient predicted by a model at an angle of attack $2^\circ$ past stall since this is late enough to substantially penalize the prediction of attached flow if the experimental flow is separated, but not so late that the experimental three-dimensionality challenges the modeling assumptions of 2D steady RANS, making detached eddy simulation (DES) necessary. Since a DES model is constructed as an extension of a 2D steady RANS model, it has been observed that the limited capability of the baseline SST model to predict the onset of separation limits its performance in DES contexts at angles of attack near stall, but this dependence on the underlying RANS model diminishes at higher angles of attack \citep{Bidadi2023}. This suggests that the suitability of the proposed 2D steady RANS models should be determined based on their prediction near the onset of stall, and less emphasis should be placed on deficiencies at higher angles of attack. The resulting recommended coefficients for the $a_{1,\mathrm{APG}}$ model are 
\begin{align}
    \begin{split}
        s_T&=0.5, \\
        a_{1,\mathrm{APG}}&=0.265, \\
    \end{split}
\end{align}
and the resulting recommended coefficients for the $\beta^*_\mathrm{APG}$ model are 
\begin{align}
    \begin{split}
        s_T&=0.5, \\
        \beta^*_\mathrm{APG}&=0.108.
    \end{split}
\end{align}
Note that $s_T=0.5$ is recommended for both models. 

\begin{figure}
\centering
\floatsetup[figure]{style=plaintop} 
\captionsetup[subfigure]{position=top, justification=raggedright, singlelinecheck=false} 
\subcaptionbox{}{\includegraphics[width=0.49\textwidth]{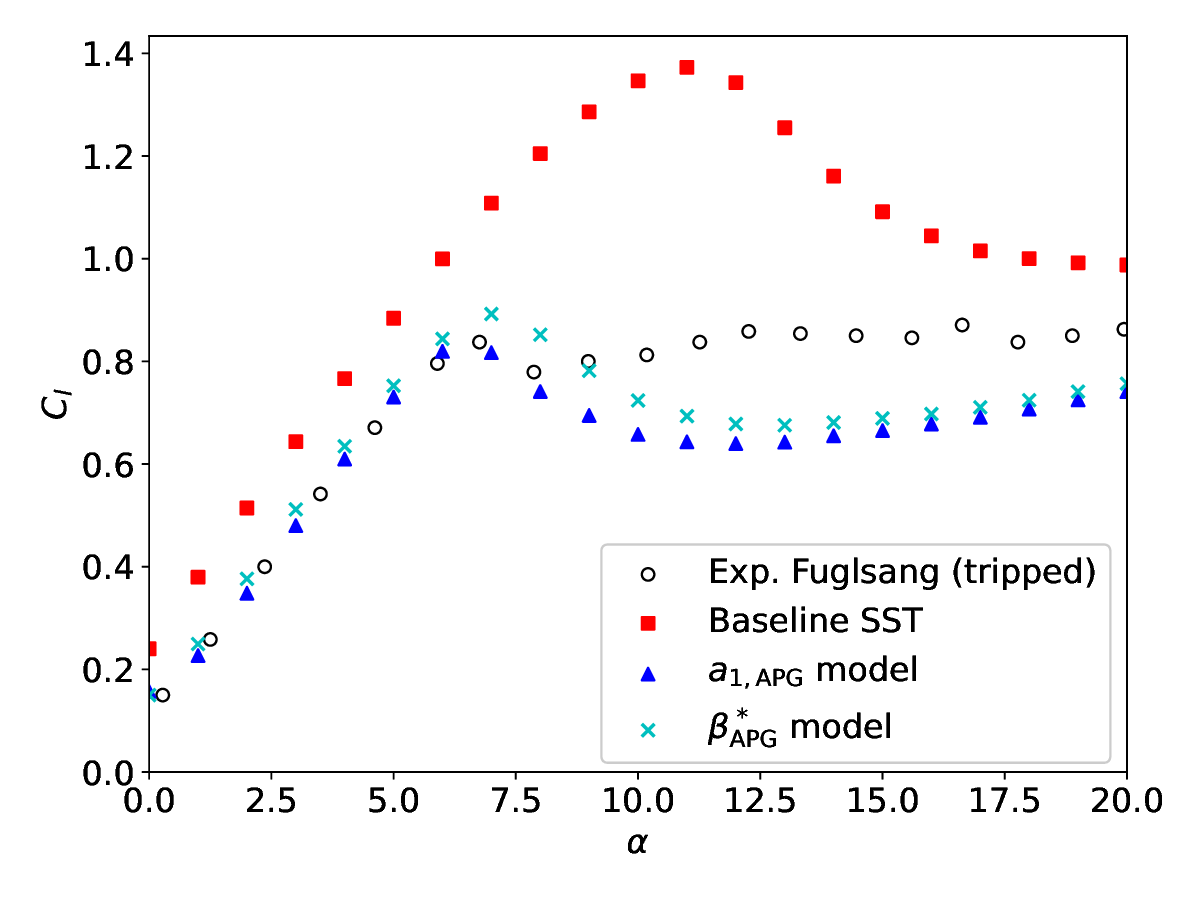}}
\caption{Lift coefficient versus angle of attack for the FFA-W3-301 airfoil at $Re_c=1.6\times10^6$. Experimental reference data \citep{Fuglsang1998} and predictions from the baseline SST RANS model and two proposed RANS models using Nalu-Wind.}
\label{fig:ffa}
\end{figure}

In Fig.~\ref{fig:ffa}, the predictions of lift coefficient are plotted versus angle of attack for the FFA-W3-301 airfoil at $Re_c = 1.6 \times 10^6$ including predictions of the baseline SST model, that of the two proposed models, and the experimental data of \cite{Fuglsang1998}.
The first observation is that the proposed models predict stall (the sudden change in slope of the lift curve) at approximately the same angle of attack as in the experiment. Meanwhile, the baseline SST model predicts the stall around $5^\circ$ too late. The physical mechanism for this improvement in the lift prediction is the accurate prediction of a trailing-edge separation bubble on the suction side of the airfoil, and detailed evidence will be discussed below.

The second observation of Fig.~\ref{fig:ffa} is that the proposed models improve the prediction of the lift in the linear regime ($\alpha \lessapprox 7^\circ$) while the baseline SST overpredicts the lift in this regime. The error of the baseline SST in the linear regime is particularly striking (although consistent with XFOIL and EllipSys2D calculations \citep{Fuglsang1998} for this airfoil) because the linear regime of airfoils is typically accurately predicted for thin airfoils. The APG sensor in the proposed models is clearly activated in these flows since the predictions differ from the baseline SST model, and for both proposed models, their treatment of the APG region is improving the prediction of lift. The physical mechanism for this improvement is that the relative thickness of this airfoil leads to sufficiently strong APG, even at low angles of attack, to trigger a pressure-side separation, and this is captured by the models as will be discussed below.

The third observation of Fig.~\ref{fig:ffa} is that for $\alpha \ge 10$, we observe that the proposed models underpredict the lift. This is acceptable since the experimental flow is likely highly three-dimensional in this regime (a few degrees past stall) due to wind tunnel effects (c.f., \cite{Gleyzes2003}). Accurate prediction in this regime often requires spatial- and temporal-scale resolving simulations (e.g., DES) since RANS eddy viscosity models typically do not accurately predict the turbulence in a separation bubble. However, DES struggles with the prediction of the onset of turbulence, since the onset typically is dictated by the underlying RANS model \citep{Bidadi2023}.

The fourth observation on Fig.~\ref{fig:ffa} is that the two proposed models have similar performance across the full range of angles of attack explored. This suggests that the aspect that the models have in common, i.e., our proposed adverse pressure gradient sensor in Eq.~\ref{proposed:sst_apg}, is perhaps more consequential than the different ways in which the models modify the eddy viscosity within the APG region (through Eqs. \ref{eq:mut_mod} vs. \ref{eq:k_mod}). 

\begin{figure}
\centering
\floatsetup[figure]{style=plaintop} 
\captionsetup[subfigure]{position=top, justification=raggedright, singlelinecheck=false} 
\subcaptionbox{}{\includegraphics[width=0.4\textwidth]{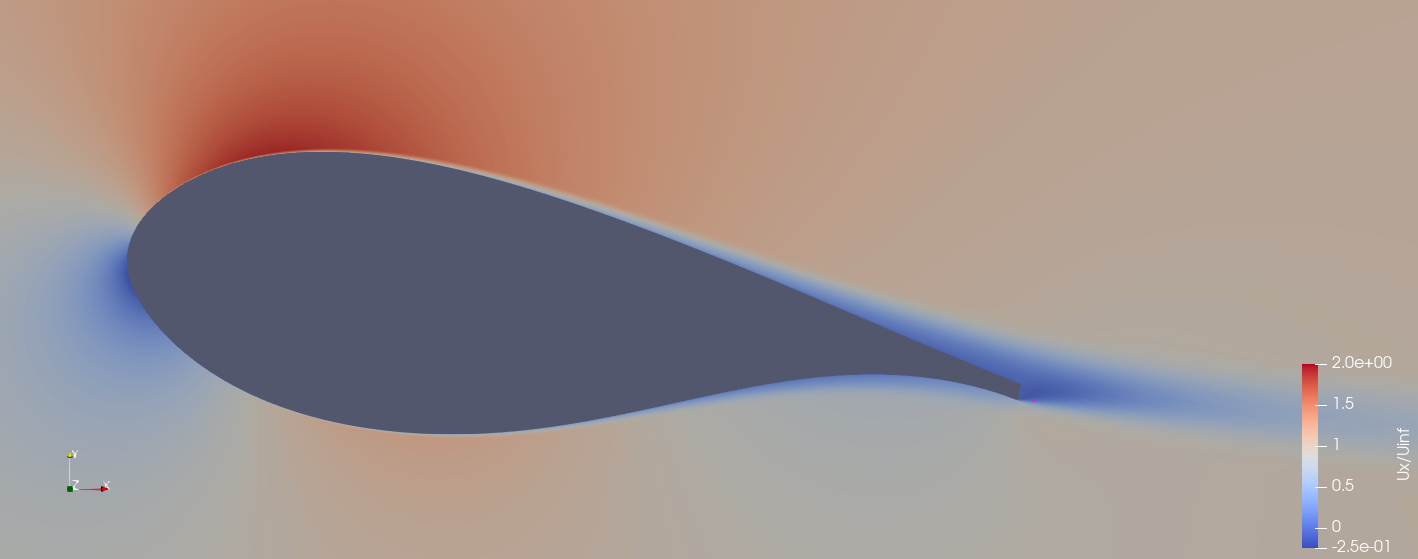}} \\
\subcaptionbox{}{\includegraphics[width=0.4\textwidth]{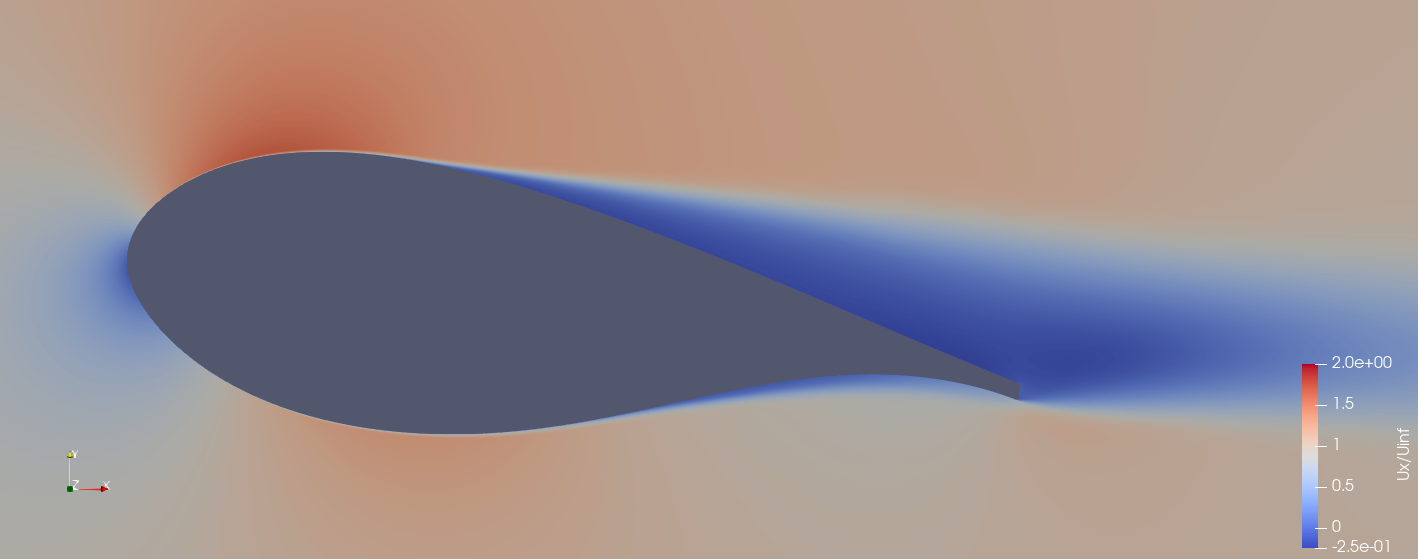}} \\
\subcaptionbox{}{\includegraphics[width=0.4\textwidth]{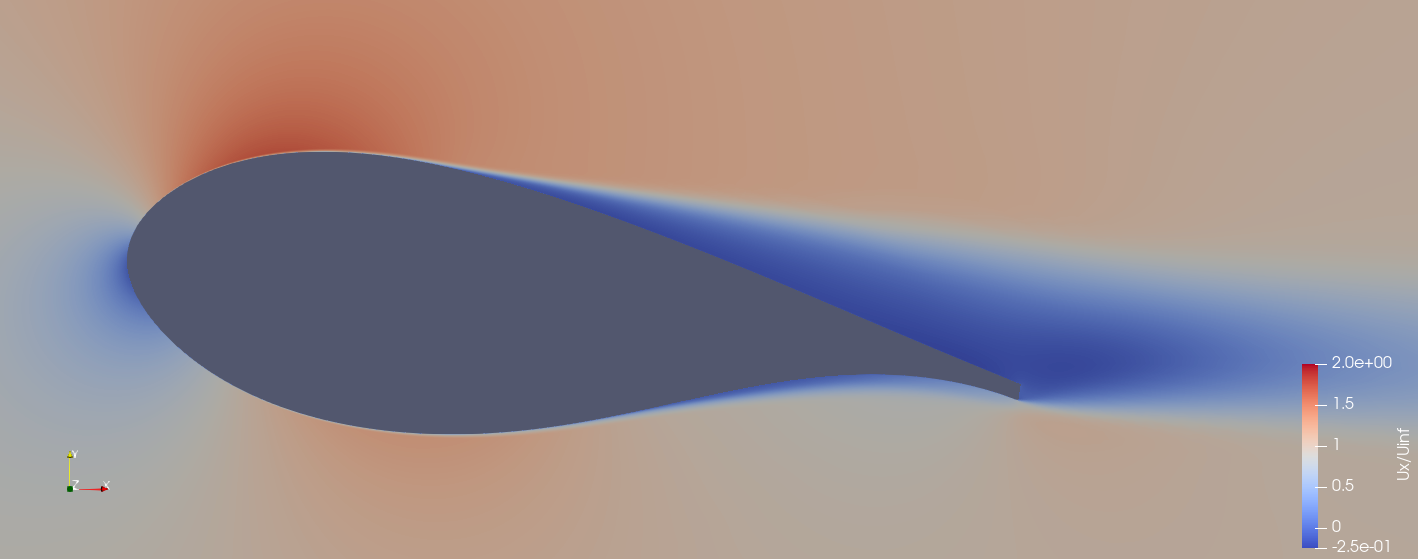}}
\caption{Contours of the streamwise component of velocity $U_x/U_\infty$ for the baseline SST (a), $a_{1,\mathrm{APG}}$ (b), and $\beta_\mathrm{APG}^*$ (c) models for the FFA-W3-301 airfoil at $Re_c=1.6\times10^6$ and $\alpha=9^\circ$.}
\label{fig:vis_aoa9}
\end{figure}

\begin{figure}
\centering
\floatsetup[figure]{style=plaintop} 
\captionsetup[subfigure]{position=top, justification=raggedright, singlelinecheck=false} 
\subcaptionbox{}{\includegraphics[width=0.4\textwidth]{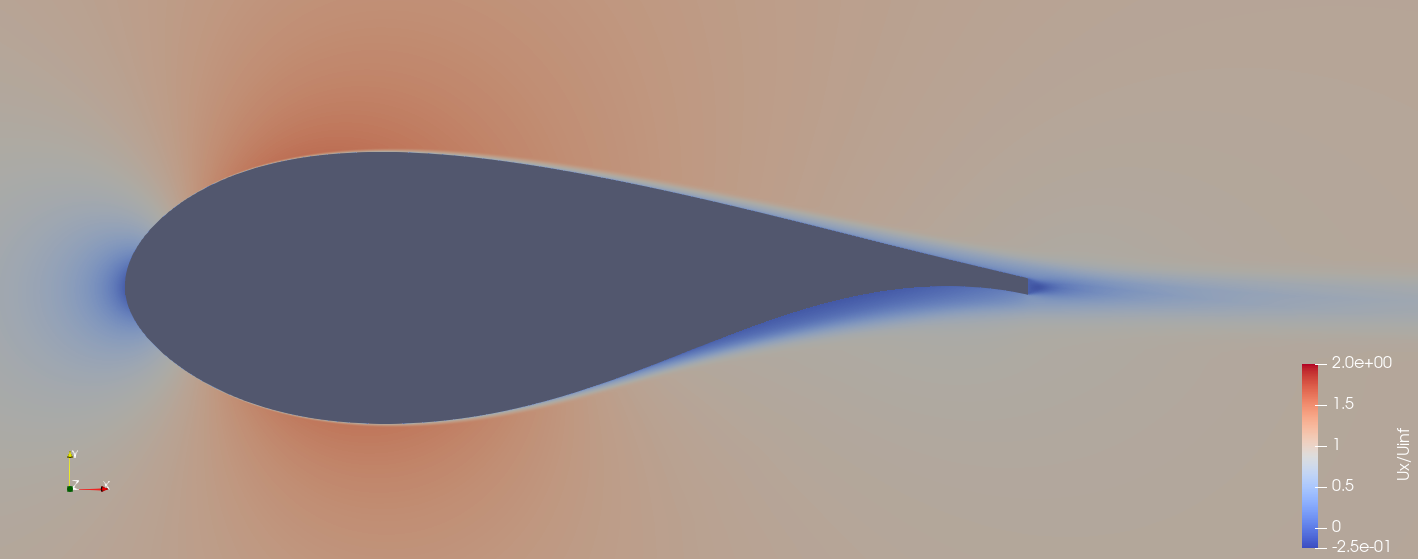}} \\
\subcaptionbox{}{\includegraphics[width=0.4\textwidth]{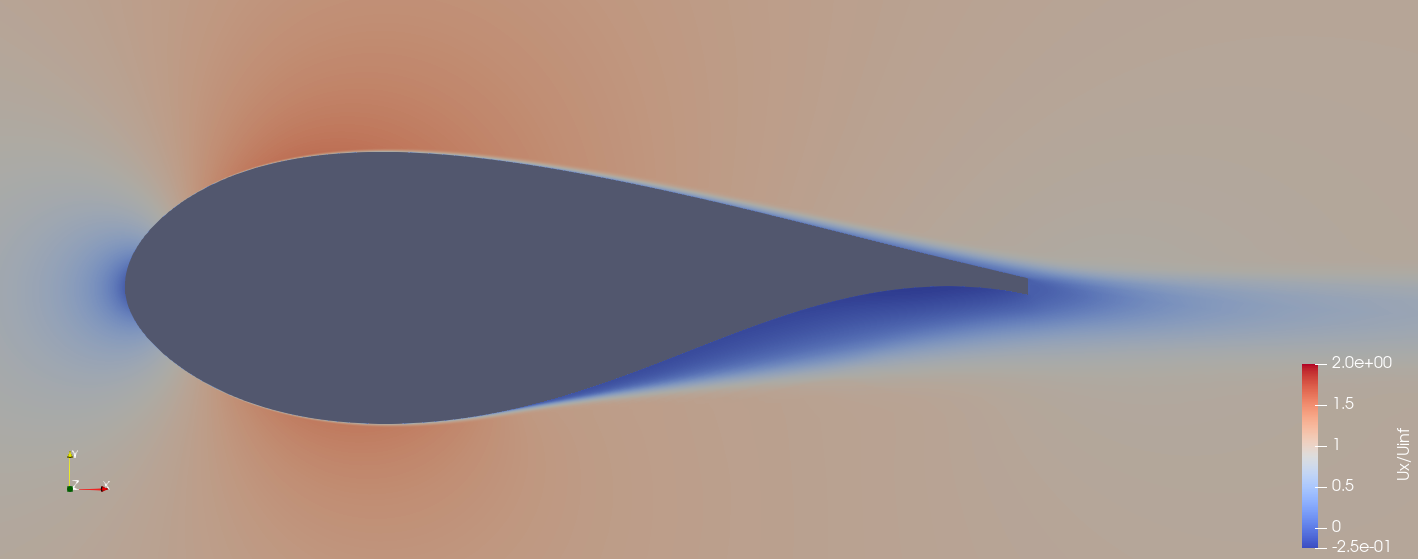}} \\
\subcaptionbox{}{\includegraphics[width=0.4\textwidth]{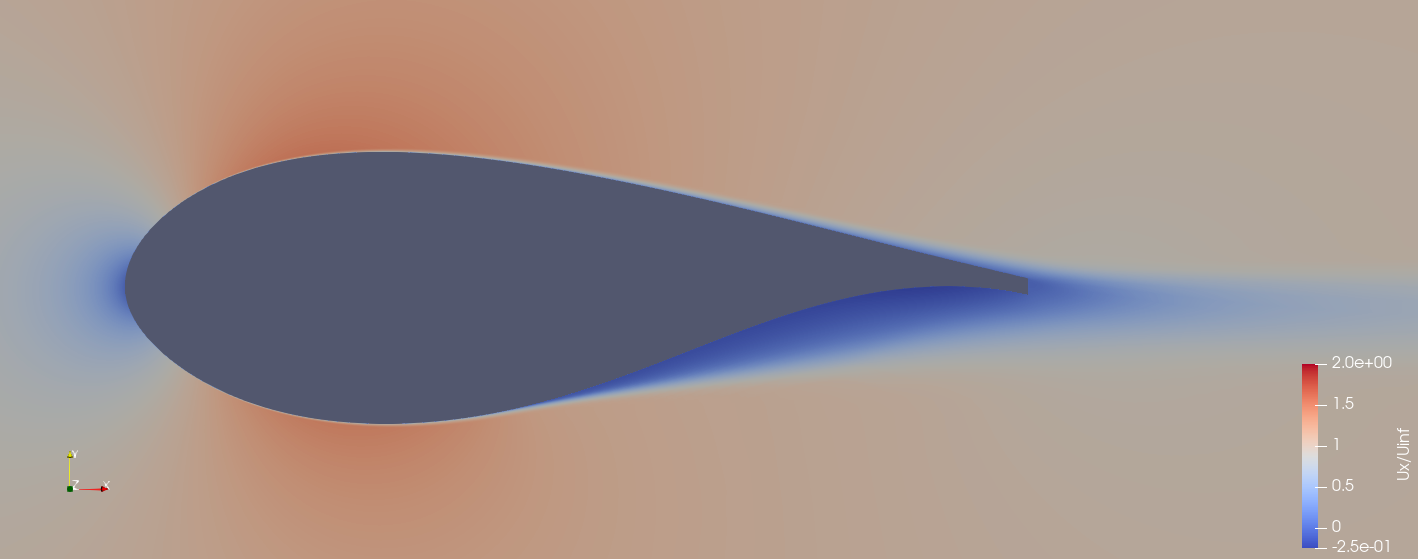}}
\caption{Contours of the streamwise component of velocity $U_x/U_\infty$ for the baseline SST (a), $a_{1,\mathrm{APG}}$ (b), and $\beta_\mathrm{APG}^*$ (c) models for the FFA-W3-301 airfoil at $Re_c=1.6\times10^6$ and $\alpha=0^\circ$.}
\label{fig:vis_aoa0}
\end{figure}

Contours of the streamwise component of velocity $U_x/U_\infty$ for the three models are analyzed at two angles of attack to visually display the extent of the separations predicted by each of the models and to help contextualize the quantitative lift predictions discussed above.
In Fig.~\ref{fig:vis_aoa9}, the results are presented for $\alpha=9^\circ$, which is experimentally observed to be past the stall angle. The proposed models predict large separation bubbles on the suction side of the airfoil, which is consistent with their accurate characterization of stall (see Fig.~\ref{fig:ffa}). Meanwhile, the baseline model predicts a mostly attached flow at this angle of attack (with perhaps a small separation at the blunt trailing edge), which is consistent with the model's overprediction of lift at this angle of attack (see Fig.~\ref{fig:ffa}).
In Fig.~\ref{fig:vis_aoa0}, the results for $\alpha=0^\circ$ are presented. Similarly, the baseline SST predicts a nearly attached flow; the proposed models predict a separation on the pressure side of the airfoil. Since the baseline model overpredicts lift in the linear regime of the lift curve (see Fig.~\ref{fig:ffa}) and the proposed models accurately predict lift in this regime, it appears that the proposed models' prediction of the pressure-side separation explains their improved agreement with experimental lift data in the linear regime.


\begin{figure}
\centering
\floatsetup[figure]{style=plaintop} 
\captionsetup[subfigure]{position=top, justification=raggedright, singlelinecheck=false} 
\subcaptionbox{}{\includegraphics[width=0.49\textwidth]{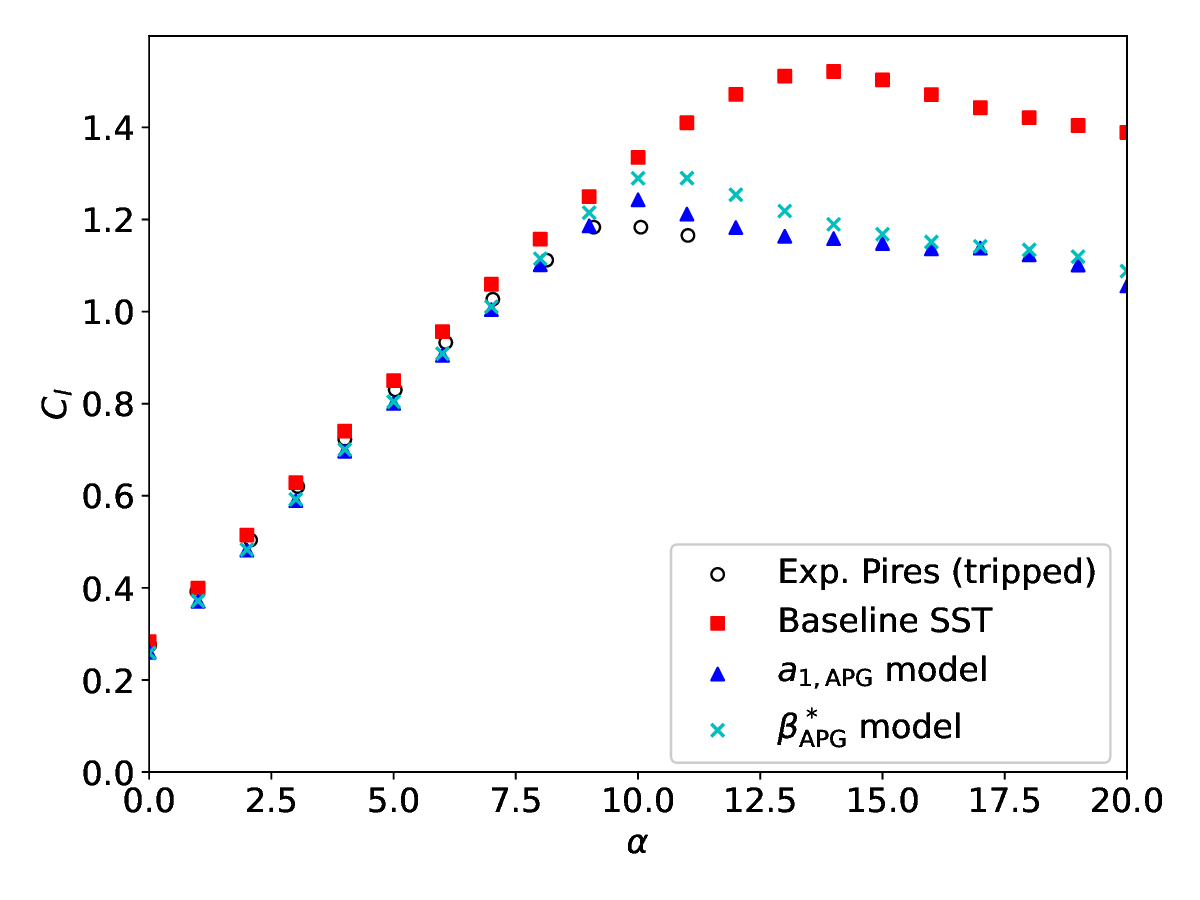}}
\subcaptionbox{}{\includegraphics[width=0.49\textwidth]{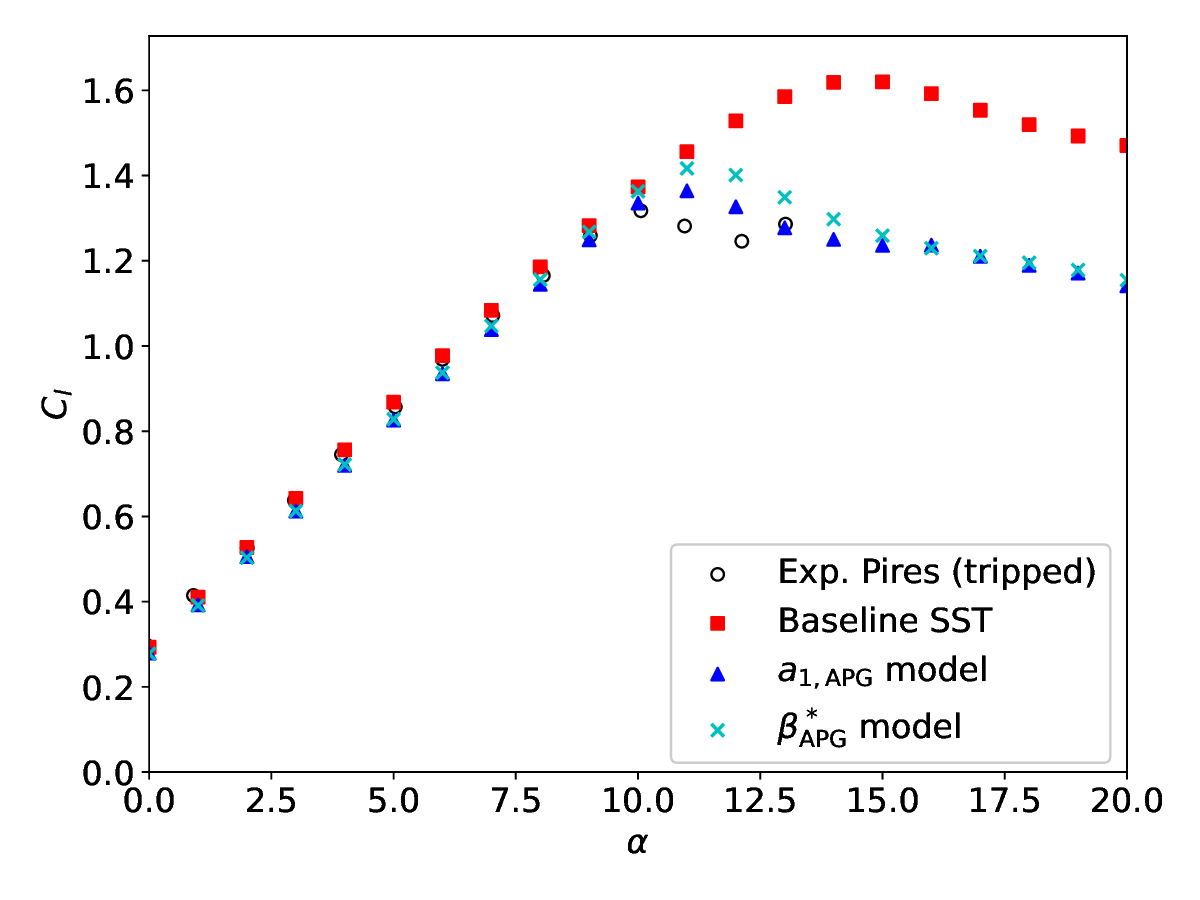}} \\
\subcaptionbox{}{\includegraphics[width=0.49\textwidth]{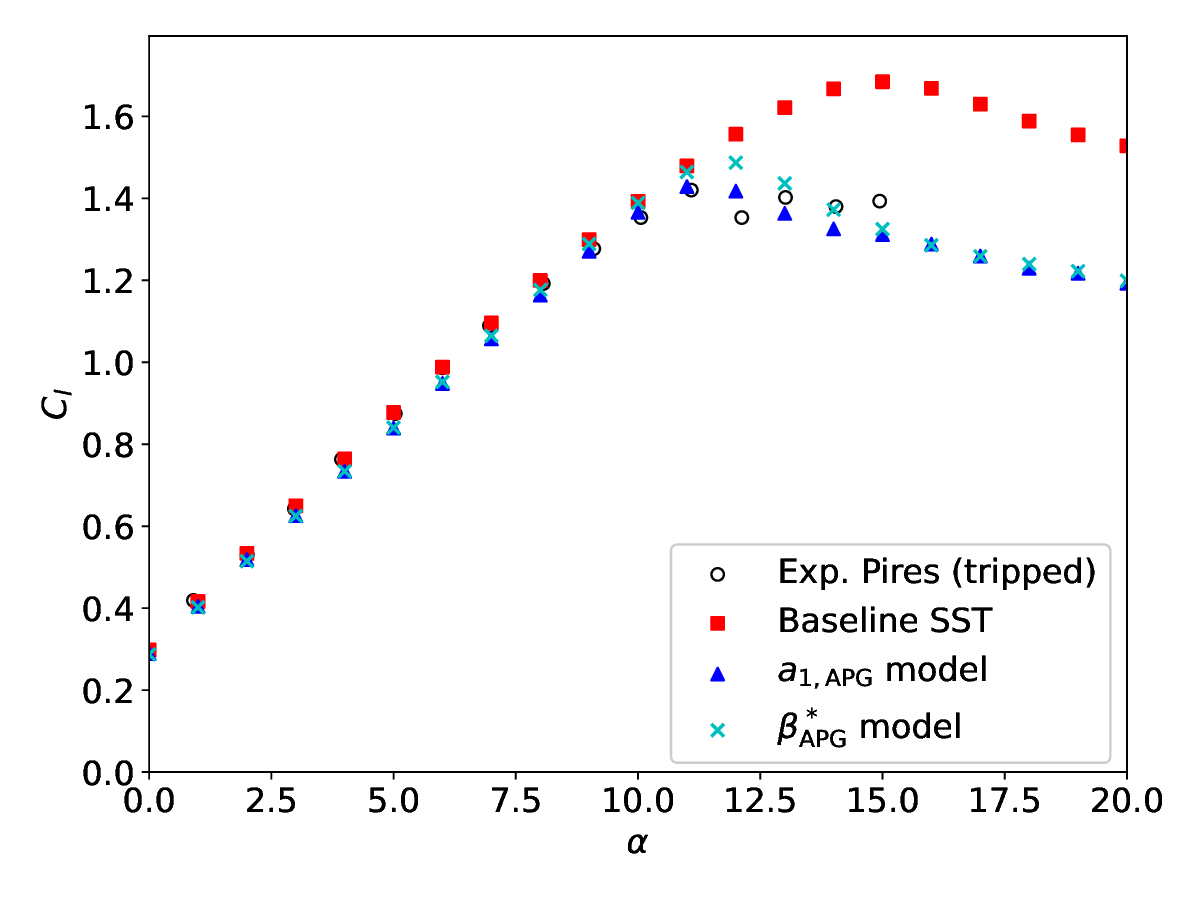}}
\subcaptionbox{}{\includegraphics[width=0.49\textwidth]{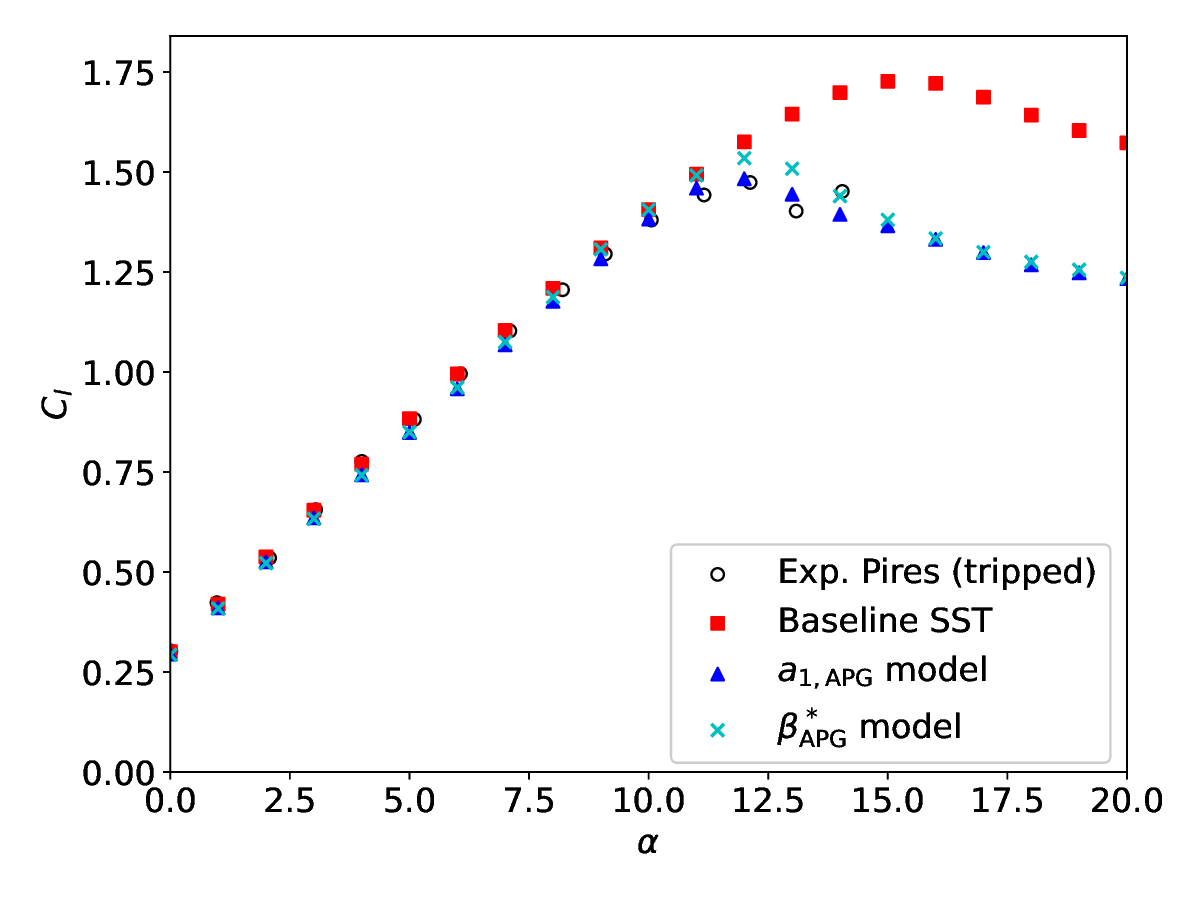}} \\
\subcaptionbox{}{\includegraphics[width=0.49\textwidth]{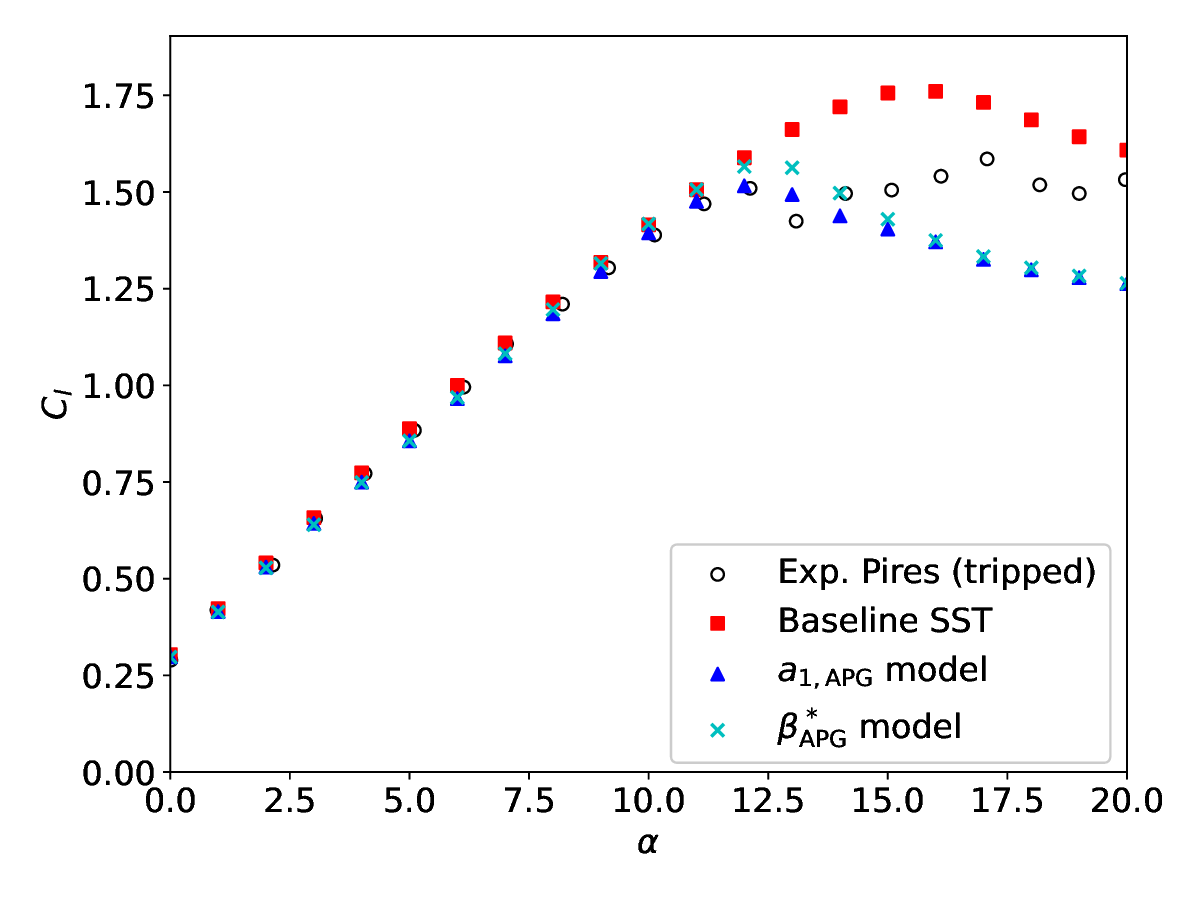}}
\caption{Lift coefficient versus angle of attack for the DU00-W-212 airfoil at $Re_c=(3, 6, 9, 12, 15)\times10^6$ in panels a, b, c, d, and e, respectively. Experimental reference data \citep{Pires2016} and predictions from the baseline SST RANS model and two proposed RANS models using Nalu-Wind.}
\label{fig:du_Re}
\end{figure}

To analyze the ability of the models to generalize across geometries and Reynolds numbers, we consider the experimental investigation of the DU00-W-212 airfoil by \cite{Pires2016} for a range of $Re_c$ from $3 \times 10^6$ to $1.5 \times 10^7$. In Fig.~\ref{fig:du_Re}, the lift predictions are presented versus angle of attack. Across all Reynolds numbers, the proposed models predict the onset of stall much more accurately than the baseline SST model does. For higher angles of attack past stall, the models underpredict the lift, but as remarked above, this is not a point of focus since 2D steady RANS is not well motivated in this regime. For this airfoil, all models, including the baseline SST, predict the linear regime well. Accurate prediction in the linear regime is credited to the robust performance of the baseline SST model when applied to boundary layers with modest pressure gradients. The fact that the predictions of the proposed and baseline models agree indicates that pressure gradients for this airfoil are not strong enough, at low angles of attack, to activate the proposed pressure gradient sensor. Thus, for this airfoil, at low angles of attack, the proposed model recovers the SST model, which is accurate in this regime. Comparing the linear regime of the DU00-W-212 to that of the thicker FFA-W3-301 airfoil, the proposed sensor remains inactive in the former case and the baseline model is accurate, and in the latter case, the  sensor activates to correct the inaccurate baseline model. This is expected since the thicker airfoil leads to greater streamline curvature and thus stronger pressure gradients.

\begin{figure} 
\centering 
\includegraphics[width=0.6\textwidth]{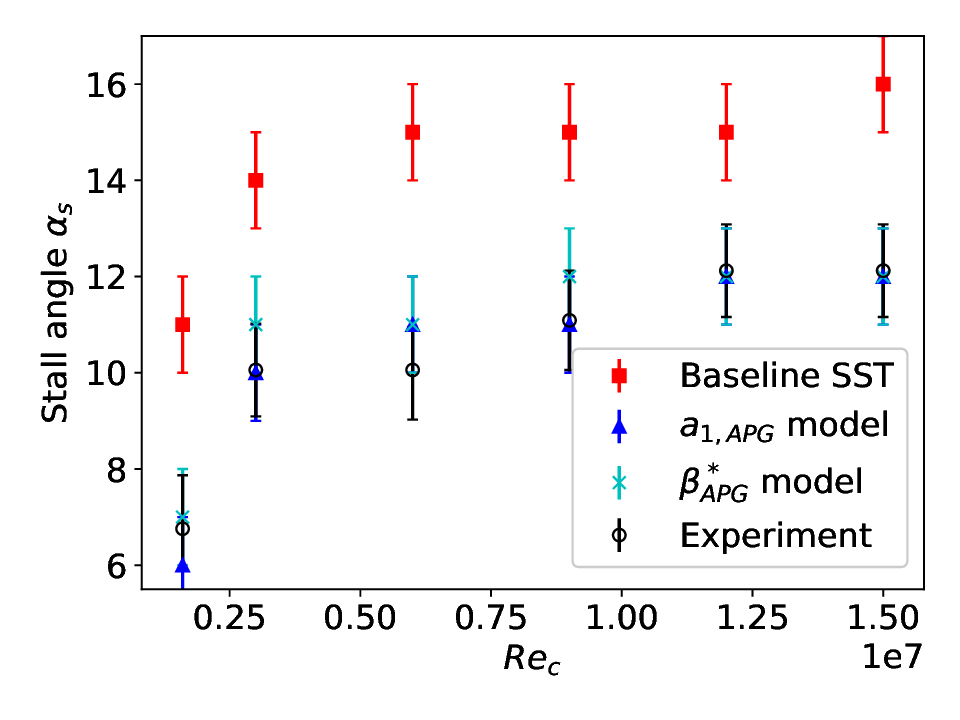}
\caption{The stall angle versus the chord-based Reynolds number for the FFA-W3-301 airfoil ($Re_c=1.6\times10^6$) and DU00-W-212 airfoil (otherwise). Experimental reference data \citep{Fuglsang1998,Pires2016} and predictions from the baseline SST RANS model and two proposed RANS models using Nalu-Wind.}
\label{fig:stall_angle}
\end{figure}

Next, we reinterpret the prior results by focusing on the angle of attack at which stall occurs as this is of particular interest for engineering design. For the purpose of the analysis, we define the stall angle $\alpha_s$ as the lowest angle of attack at which $C_{L,\mathrm{max}}$, the local maximum in $C_L$ vs. $\alpha$, occurs. 
In Fig.~\ref{fig:stall_angle}, $\alpha_s$ is plotted for the chord-based Reynolds number $Re_c$ for each of the cases previously considered, i.e., the FFA-W3-301 airfoil at $Re_c=1.6\times10^6$ and the DU00-W-212 airfoil at $Re_c=(3, 6, 9, 12, 15)\times10^6$. Error bars plotted are epistemic uncertainty intervals computed by measuring the difference in the angle of attack between the stall angle identified and the data at the next highest and lowest angles of attack measured. Since the simulations are run at $1^\circ$ intervals, the error bars are $+/-1^\circ$. For the experiment, they are slightly different based on the angles of attack reported in the experimental data.
It is observed that the proposed models lead to predictions of the stall angle that are accurate across all Reynolds numbers and airfoils investigated. Indeed, the predictions of the proposed models improve the prediction of the stall angle by three to five degrees with respect to baseline SST model for all conditions considered. 
It is important for models to accurately predict $\alpha_s$, as it determines the range of angles of attack that the airfoil can operate while avoiding stall. The large error of the baseline SST model in predicting $\alpha_s$ would necessitate conservative safety margins, limiting performance, to avoid stall during operation.

\begin{figure} 
\centering 
\includegraphics[width=0.6\textwidth]{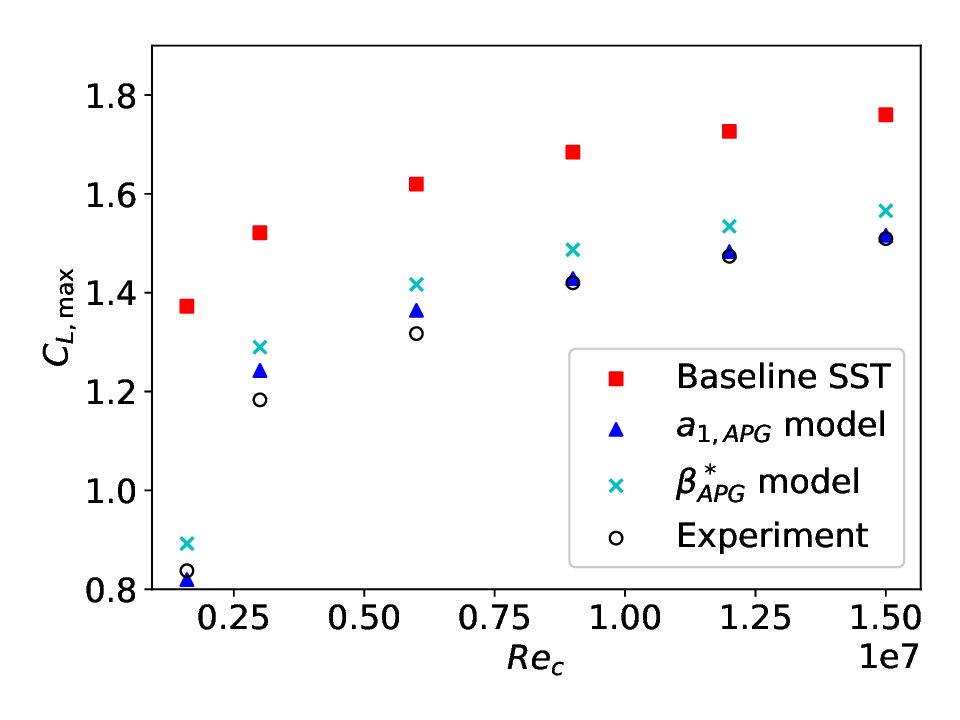}
\caption{The maximum lift coefficient versus the chord-based Reynolds number for the FFA-W3-301 airfoil ($Re_c=1.6\times10^6$) and DU00-W-212 airfoil (otherwise). Experimental reference data \citep{Fuglsang1998,Pires2016} and predictions from the baseline SST RANS model and two proposed RANS models using Nalu-Wind.}
\label{fig:max_cl}
\end{figure}

The prediction of the lift produced at the stall angle is reported in Fig.~\ref{fig:max_cl} for the same cases considered in Fig.~$\ref{fig:stall_angle}$. The baseline SST model considerably overpredicts $C_{L,\mathrm{max}}$, especially at low Reynolds numbers. Meanwhile, the proposed models provide relatively accurate predictions, with the $a_{1,\mathrm{APG}}$ model slightly outperforming the $\beta_\mathrm{APG}^*$ model.
The accurate prediction of $C_{L,\mathrm{max}}$ is important for engineering design, as it determines peak aerodynamic loads. Consider the example of the design of a wind turbine; $C_{L,\mathrm{max}}$ determines the maximum power generated, which determines the peak stresses on the mechanical, structural, and electrical systems. Underpredicting this quantity could lead to system failure. Overpredicting would lead to a more expensive turbine that cannot realize its nominal peak performance.

In summary, the model trained on the FFA-W3-301 airfoil has generalized well to the DU00-W-212 airfoil. This demonstrates insensitivity to moderate changes in geometry because the former has a maximum thickness of 30\% while the latter has a maximum thickness of 20\%. Furthermore, the model has generalized well across almost a decade of Reynolds numbers from $Re_c=1.6 \times 10^6$ to up to $Re_c=1.5\times 10^7$.

\subsection{Boeing Gaussian bump results}
The models are applied to the Boeing Gaussian bump to assess the effect of more substantial geometric changes and to assess the prediction of viscous drag. 
Several recent studies \citep{Balin2020a,Iyer2021,Prakash2022,Whitmore2021,Zhou2023} in turbulence model validation and development have, or have almost, exclusively focused on the Boeing Gaussian bump, a challenging geometry for which satisfactory predictions with RANS have yet to be reported. The flow includes a nominally zero-pressure-gradient upstream section, smooth-body separation, and smooth-body reattachment, which include the key building blocks for external aerodynamics; this indeed was the goal in the design of this CFD validation experiment Williams et al. \citep{Williams2020}.

\begin{figure}
\centering
\floatsetup[figure]{style=plaintop} 
\captionsetup[subfigure]{position=top, justification=raggedright, singlelinecheck=false} 
\subcaptionbox{}{\includegraphics[width=0.49\textwidth]{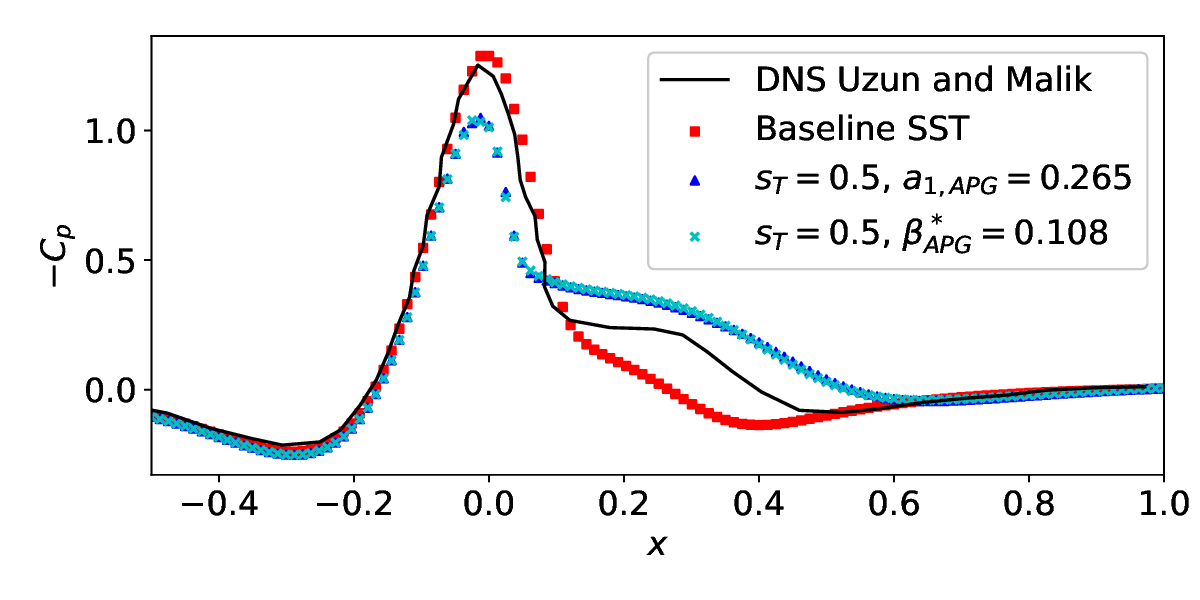}}
\subcaptionbox{}{\includegraphics[width=0.49\textwidth]{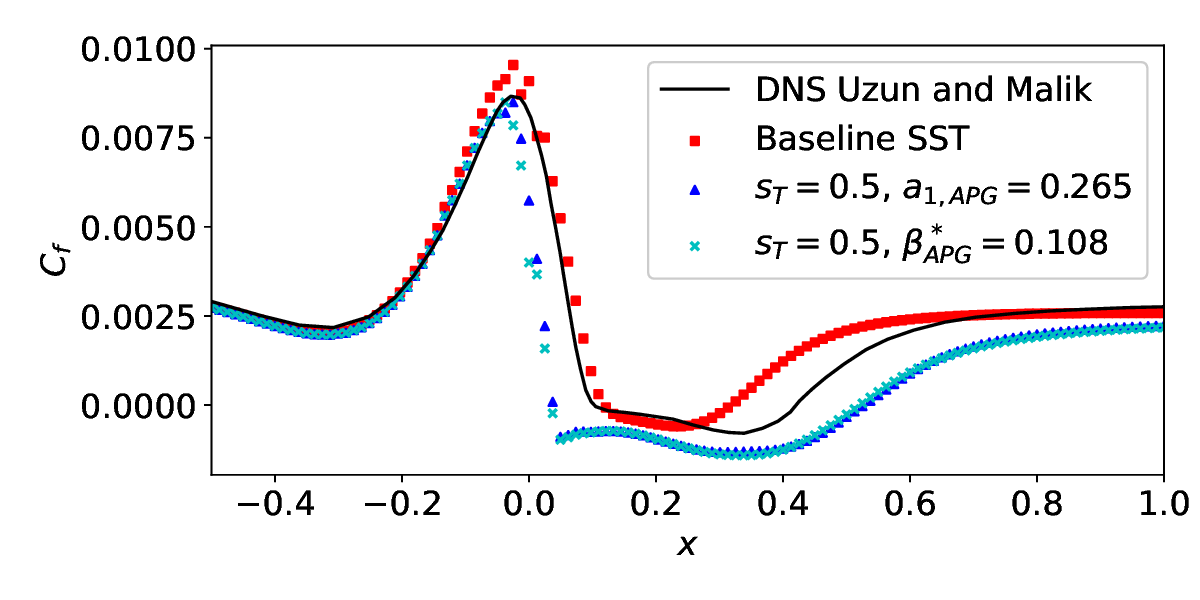}} \\
\subcaptionbox{}{\includegraphics[width=0.49\textwidth]{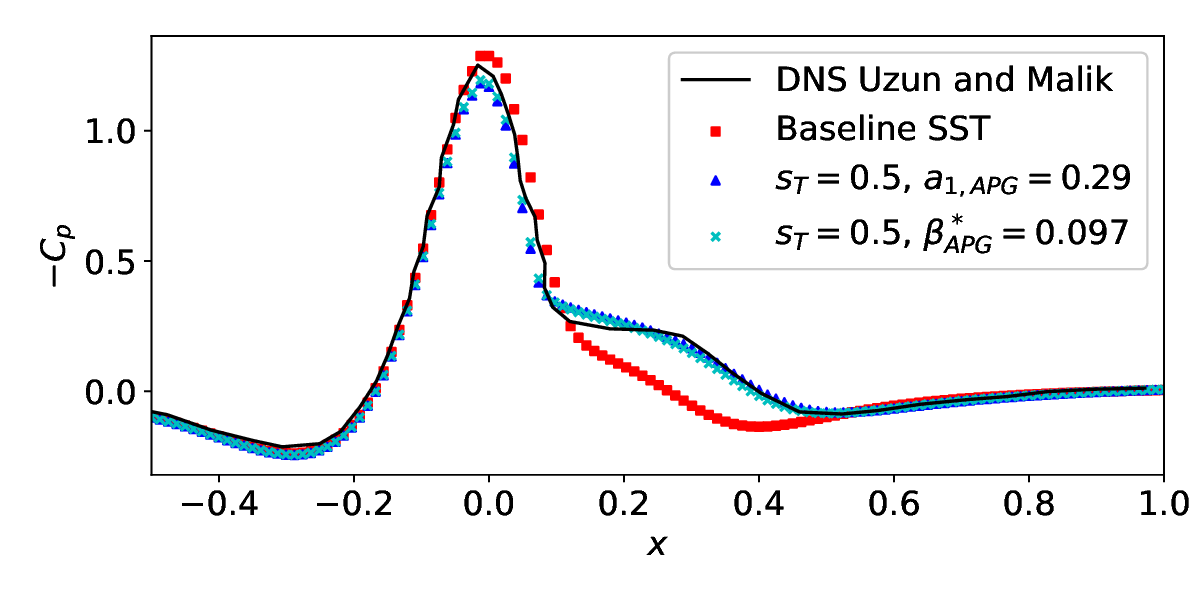}}
\subcaptionbox{}{\includegraphics[width=0.49\textwidth]{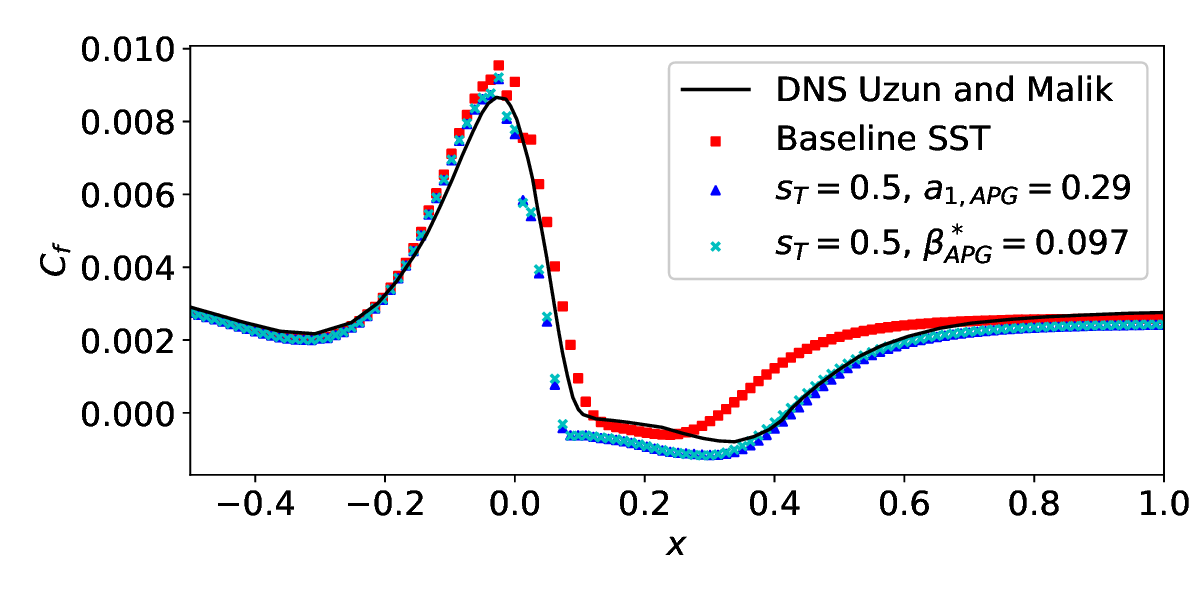}} \\
\subcaptionbox{}{\includegraphics[width=0.49\textwidth]{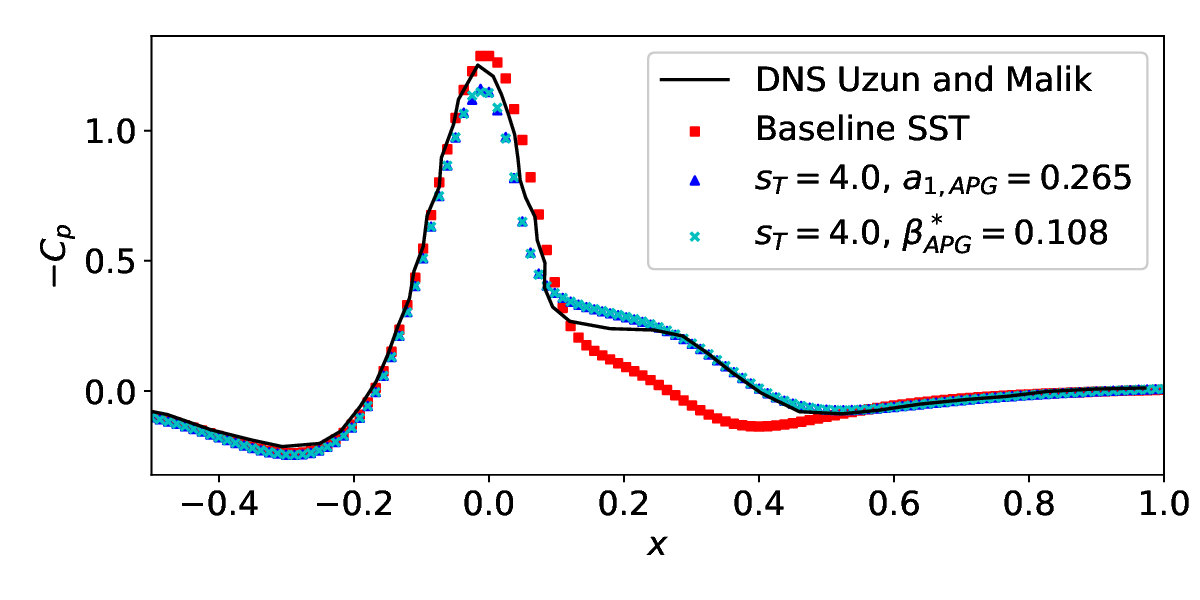}}
\subcaptionbox{}{\includegraphics[width=0.49\textwidth]{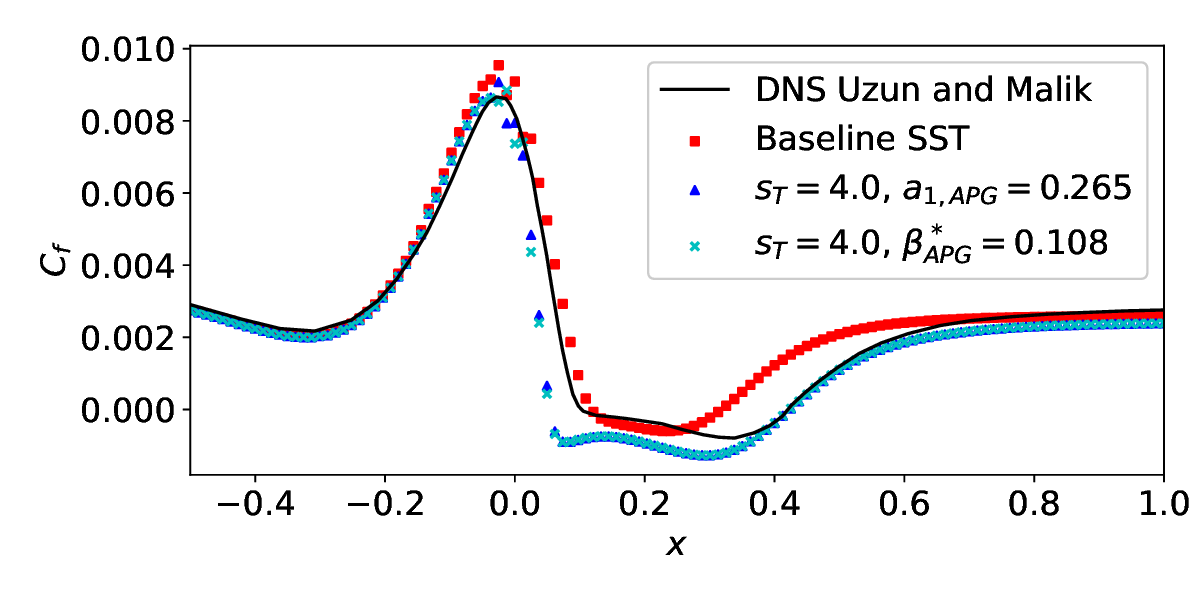}}
\caption{Coefficients of pressure (a,c,e) and friction (b,d,f) on the Boeing Gaussian bump. DNS data \citep{Uzun2022} and Nalu-Wind simulations with the baseline SST,~$a_{1,\mathrm{APG}}$,~and $\beta^*_\mathrm{APG}$ models with various choices of model coefficients.}
\label{fig:bump_results}
\end{figure}

In Fig.~\ref{fig:bump_results}, the predictions of the coefficients of pressure and friction are plotted versus the streamwise coordinate for various choices of the model coefficients. Figure \ref{fig:bump_results}a and b indicate the results for the model coefficients proposed previously, which provide accurate predictions of the onset of separation for two airfoils. For both models, these sets of coefficients leads to over-separated flows as indicated by the enlarged region of negative $C_f$ with respect to that of the DNS data. 
Figure  \ref{fig:bump_results}c and d indicate the result of holding $s_T=0.5$ and recalibrating $a_{1,\mathrm{APG}}$ and $\beta^*_\mathrm{APG}$ via a grid search (procedure described previously) to optimize the agreement with the DNS data for the bump, which indeed leads to an accurate prediction. Figure \ref{fig:bump_results}e and f indicate the result of holding $a_{1,\mathrm{APG}}$ and $\beta^*_\mathrm{APG}$ to the values from the prior section and optimizing $s_T$ via a grid search. This also leads to good agreement with the DNS data. These approaches indicate that there are two paths to building a more universal model: (1) developing variable models for $a_{1,\mathrm{APG}}$ and $\beta^*_\mathrm{APG}$ or (2) developing a variable model for $s_T$ (or a more general nondimensionalization of the pressure gradient).

Recall our second modeling objective is to not degrade the performance of the baseline SST in regions where it is performing well, such as in most attached flows. For the Boeing Gaussian bump, the baseline SST performs well on the upstream half of the bump ($x\lessapprox 0$). Also, recall that for ad hoc tuning of $a_1$ and $\beta^*$ as shown in Figs. \ref{fig:adhoc_a1} and \ref{fig:adhoc_betas}, the modified coefficients degrade the skin friction predictions for $x<0$. Meanwhile, for the presently proposed models, for all three choices of coefficients presented in Fig.~\ref{fig:bump_results}, the models preserve the accurate skin friction predictions for $x\lessapprox0$, indicating that our second modeling objective is achieved. Indeed, Fig. \ref{fig:apg}e and Fig. \ref{fig:sT} indicate that as long as $s_T\ge0.5$, the proposed pressure gradient sensor will prevent the modified eddy viscosity models from activating for $x\lessapprox0$ for this case.

Between $x=0$ (the apex of the bump) and the separation point (the first zero crossing of $C_f$), the predictions of the proposed models depart from that of the baseline model. The departure is explained by Fig. \ref{fig:apg}e and \ref{fig:sT} which indicates that, for $s_T$ between 0.5 and 4.0, the APG sensor is active between $x=0$ and the separation point. For the first choice of model coefficients in Fig.~\ref{fig:bump_results}a and b, the flow is over separated. This affects the pressure coefficient upstream of the separation point through subsonic ellipticity. Specifically, a larger separation bubble reduces streamline curvature and results in weaker pressure gradients, thus a reduced suction peak at $x=0$. The early drop in $C_f$ indicates that the separation bubble begins earlier, which is related to the fact that the flow is over separated (the flow also reattaches late). The fact that the flow is over separated indicates that either the modified coefficients $a_{1,\mathrm{APG}}$ and $\beta^*_\mathrm{APG}$ should be closer to their baseline values of $a_1$ and $\beta^*$ or the modified coefficients should be applied over a reduced region of the domain. These two cases are what is considered with the second and third choices of coefficients in Fig.~\ref{fig:bump_results}c,d and e,f, respectively. For both of these cases, the separation bubble is beginning early, particularly in the third case, similar as in the first case. This is indicated by the early crossing of zero of $C_f$ and the underprediction of $C_f$ between $x=0$ and the separation point. Meanwhile, the location of the reattachment point for cases two and three is accurately predicted unlike in case one. As a result the overall size of the bubble is only slightly overpredicted and the suction peak of the coefficient of pressure (at $x=0$) is only slightly below the experimental data. Due to the approximate correct size of the separation bubble, the streamline curvature and thus pressure coefficient distribution remains accurate throughout the domain.

\section{Conclusion} \label{sec:conclusion}
We identified that the adverse pressure gradient sensor used by the baseline SST turbulence model is not serving its purpose well. We proposed a simple sensor based on the dot product of the pressure gradient with the velocity unit vector. Rather than replace the existing APG sensor in the baseline SST model, we add our sensor on top of the baseline model to preserve the model's performance when our sensor is inactive, and we propose two interventions when our sensor is active (in strong APG regions). The first intervention we refer to as the $a_{1,\mathrm{APG}}$ model, which reduces the value of the eddy viscosity coefficient in APG regions. The second intervention we refer to as the $\beta^*_\mathrm{APG}$ model, which increases the kinetic energy destruction coefficient in APG regions. Both of these interventions are physically motivated to promote separation by directly or indirectly decreasing the eddy viscosity in regions of strong adverse pressure gradients, leading to reduced diffusion of momentum from the boundary layer edge to the wall and thus more readily permitting near-wall flow reversal. Coefficient values are calibrated with \textit{a posteriori} simulation data. In addition, we calibrate the APG threshold constant $s_T$, which determines how strong the APG must be to be labeled as an APG region.

Our two proposed models satisfy our modeling objectives, the first of which is that the models improve the prediction of boundary layer separation. This is demonstrated on the Boeing Gaussian bump and on two thick airfoils, the FFA-W3-301 and the DU00-W-212 airfoils, which are representative of the cross section of modern wind turbine blades. While the baseline SST predicts stall $3^\circ$ to $5^\circ$ late for all cases considered, the proposed models predict stall within the margins of experimental uncertainty. This allows the proposed models to predict the maximum of the lift coefficient significantly more accurately than the baseline SST model, which has important implications for aircraft and wind turbine design. Both models also improve the prediction of lift in the linear regime of the 30\% thick FFA-W3-301 airfoil likely due to their improved characterization of a pressure-side separation bubble. Our two proposed models have similar performance (despite fairly different mechanisms for separation enhancement), which indicates that the proposed adverse pressure gradient sensor, a common feature of both models, is likely the reason for their success.

Our second modeling objective is that the models should not worsen the performance of the SST model in attached flows. In the Boeing Gaussian bump, for choices of the APG sensor threshold coefficient $s_T \ge 0.5$, the model performance outside of the separated region is not affected by the choice of model coefficients. This isolation of the modeling choices is particularly important for the preservation of the accurate upstream prediction of the skin friction coefficient by the baseline SST model, which is inherited by the proposed models by construction.

Our third modeling objective is that model coefficients should not change with modest changes in Reynolds number (e.g., a factor of 10) or modest changes in geometry (e.g., across different airfoil shapes). The model coefficients are held fixed for the two airfoil geometries considered. The FFA-W3-301 and DU00-W-212 airfoils have maximum thicknesses of 30\% and 20\% of their chord lengths, respectively, indicating marked geometric differences. The model coefficients are calibrated for the flow over the FFA airfoil at a Reynolds number of 1.6 million. (The coefficient values attained are $s_T=0.5$, $a_{1,\mathrm{APG}}=0.265$, and $\beta^*_\mathrm{APG}=0.108$.) The models are tested with these coefficients held fixed on the DU airfoil at Reynolds numbers between 3 and 15 million, and substantial improvements to the prediction of stall are observed for all cases. It should be noted that the models are tested at nearly 10 times higher Reynolds numbers than they were trained. Other studies have tested their models at Reynolds numbers that are the same as their training Reynolds numbers \cite{Bangga2018,Barkalov2022} or only 50\%  higher \cite{Singh2017}. The relative robustness of the model lends confidence in recommending the use of the proposed models for other airfoils at similar Reynolds numbers, which satisfies our third modeling objective. 

However, we did not develop fully general turbulence models.
Upon applying our models to the Boeing Gaussian bump, we find, that our models, like those proposed in \cite{Balin2020a,Iyer2021,Prakash2022,Whitmore2021,Zhou2023}, do not satisfactorily predict separation. However, we do demonstrate unparalleled accuracy for a RANS simulation when flow-specific calibration of the model coefficients is performed. We also demonstrate that, without tuning, the sensor that we developed identifies the regions where the baseline model breaks down, which may be useful for future studies of this flow. The study of Prakash et al. \citep{Prakash2024} observed that the traditional viscous unit non-dimensionalization of the Reynolds stress profiles fails to collapse DNS data for this flow. They propose a novel non-dimensionalization of the Reynolds stress profiles for this flow and argue that the failure of the traditional viscous non-dimensionalization explains why existing turbulence models have struggled with this flow. The theory provides a possible explanation why the present model required recalibration for this flow.

The current work has studied 2D steady flows assuming fully turbulent boundary layers. Normalization of the pressure gradient sensor in terms of local flow properties may allow the model to be applied to complex geometries in which $L$ and $U_\infty$ vary in space, but the it is not advised that the model in its present form be applied to complex geometries without consideration of these length and velocity scales and appropriate validation. The range of Reynolds numbers considered in this work spans the range relevant for applications in modern wind turbine aerodynamics ($Re_c=1.6-15$ million), but applications with significantly higher $Re_c$ should be accompanied with a suitable validation study. Application to lower Reynolds numbers or flows without forced transition (tripping) may also benefit from the incorporation of the present model with existing RANS transition models. This work has focused on the prediction of the onset of stall; integration into a DES framework is recommended for applications involving massively separated flows.
\section*{Data Availability Statement}
The data that support the findings of this study are available from the corresponding author upon reasonable request.

\section*{Funding Sources}
Funding provided in part by the U.S. Department of Energy Office of Energy Efficiency and Renewable Energy, Wind Energy Technologies Office and by the Exascale Computing Project (Grant17-SC-20SC), a collaborative effort of two U.S. Department of Energy organizations (Office of Science and the National Nuclear Security Administration) responsible for the planning and preparation of a capable exascale ecosystem, including software, applications, hardware, advanced system engineering, and early testbed platforms, in support of the nation’s exascale computing imperative.

\section*{Acknowledgments}
We would like to acknowledge insightful discussions with Niloy Gupta and Prof. Karthik Duraisamy. This work was authored by the National Renewable Energy Laboratory, operated by Alliance for Sustainable Energy, LLC, for the U.S. Department of Energy (DOE) under Contract No. DE-AC36-08GO28308. The views expressed in the article do not necessarily represent the views of the DOE or the U.S. Government. The U.S. Government retains and the publisher, by accepting the article for publication, acknowledges that the U.S. Government retains a nonexclusive, paid-up, irrevocable, worldwide license to publish or reproduce the published form of this work, or allow others to do so, for U.S. Government purposes. The research was performed using computational resources sponsored by the U.S. Department of Energy’s Office of Energy Efficiency and Renewable Energy and located at the National Renewable Energy Laboratory.

\bibliographystyle{unsrt}

\bibliography{wind,speed_bump,Airfoil_data}

\end{document}